\documentclass[preprint]{aastex}

\tighten

\newcommand{\bvec}[1]{\mbox{\boldmath $#1$}}
\newcommand{\sbvec}[1]{\mbox{\boldmath $\scriptstyle #1$}}

\newcounter{subnum}[enumi]

\newcounter{subsubnum}[subnum]

\begin{document}

\title{A Fast Gridded Method for the Estimation of the Power
Spectrum of the CMB from Interferometer Data with Application 
to the Cosmic Background Imager}

\author{S. T. Myers}
\affil{National Radio Astronomy Observatory, P.O.\ Box O, Socorro, NM 87801}
\author{C. R. Contaldi, J. R. Bond, U.-L. Pen, 
D.\ Pogosyan\altaffilmark{1}, S.\ Prunet\altaffilmark{2}}
\affil{Canadian Institute of Theoretical Astrophysics, 60 St.\ George Street,
Toronto, ON M5S 3H8, Canada}
\author{J. L. Sievers, B. S. Mason, T. J. Pearson, A. C. S. Readhead, 
M. C. Shepherd}
\affil{California Institute of Technology, 1200 E.\ California Boulevard,
Pasadena, CA 91125}
\altaffiltext{1}{Department of Physics, University of Alberta, Edmonton,
Canada}
\altaffiltext{2}{Institut d'Astrophysique de Paris, 98bis Boulevard Arago,
F 75014 Paris, France}

\begin{abstract}
We describe an algorithm for the extraction of the angular power
spectrum of an intensity field, such as the cosmic microwave
background (CMB), from interferometer data.  This new method, based on the
gridding of interferometer visibilities in the aperture plane followed
by a maximum likelihood solution for bandpowers, is much faster than
direct likelihood analysis of the visibilities, and deals with
foreground radio sources, multiple pointings, and differencing.
The gridded aperture-plane estimators are also used to construct
Wiener-filtered images using the signal and noise covariance matrices used in
the likelihood analysis.  Results are shown for simulated data.  
The method has been used to
determine the power spectrum of the cosmic microwave background from
observations with the Cosmic Background Imager, and the results are
given in companion papers.
\end{abstract}

\keywords{cosmic microwave background --- cosmology: data analysis}

\section{Introduction}\label{sec:intro}

The technique of interferometry has been widely used in radio astronomy to
image the sky using arrays of antennas.  By correlating the complex
voltage signals between pairs of antennas, the field-of-view
of a single element can be sub-divided into ``synthesized beams'' of higher
angular resolution.  
In the small-angle approximation, the interferometer forms the
Fourier transform of the sky convolved with the autocorrelation of the
aperture voltage patterns.  In standard radio interferometric data analysis,
as described for example in the text by \citet{tms} and the proceedings of the
NRAO Synthesis Imaging School \citep{synim2}, the correlations or visibilities
are inverse Fourier transformed back to the image plane. 
However, there are applications such as estimation of the angular power
spectrum of fluctuations in the cosmic microwave background (CMB) 
where it is the distribution of and correlation between visibilities
in the aperture or $uv$-plane that is of most interest.

In standard cosmological models, the CMB is assumed to be a statistically
homogeneous Gaussian random field \citep{be87}.  In this case, the spherical
harmonics of the field are independent and the statistical properties are
determined by the power spectrum $C_\ell$ where $\ell$ labels the component of
the Legendre polynomial expansion (and is roughly in inverse radians).
\citet{be87} showed that in cold dark matter inspired cosmological models,
there would be features in the CMB power spectrum that reflected critical
properties of the cosmology.  Recent detections of the first few of these
``acoustic peaks'' at $\ell<1000$ in the power spectrum 
\citep{lange,maxima1,lee,halv,ne01} have
supported the standard inflationary cosmological model with
$\Omega_{tot}\approx 1$. Measurement of the higher-$\ell$ peaks and troughs,
as well as the damping tail due to the finite thickness of the last scattering
surface, is the next observational step.  Interferometers are well-suited to
the challenge of mapping out features in the CMB power spectrum, with a
given antenna pair probing a characteristic $\ell$ proportional to 
the baseline length in units of the observing wavelength (a $100\lambda$
projected baseline corresponds to $\ell\sim 628$, see \S\ref{sec:basic}).

There are many papers in the literature on the analysis of CMB anisotropy
measurements, estimation of power spectra, and the use of interferometry
for CMB studies.
General issues for analysis of CMB datasets are discussed in
\citet{bjk98,bjk00}.  \citet{hob95} present a Bayesian method for the
analysis of CMB interferometer data, using the 3-element Cosmic Anisotropy
Telescope data as a
test case.  A description of analysis techniques for interferometric
observations from the Degree Angular Scale Interferometer (DASI)
were presented in \citet{wcdh99,wcdh00}, while \citet{halv}
report on the power spectrum results from first-season of DASI observations.
\citet{ng} discusses CMB interferometry with application to the proposed AMIBA
instrument.  \citet{hob02} have recently presented an approach similar to 
ours, and demonstrate their technique on simulated Very Small Array (VSA) 
data; a brief comparison of their algorithm with ours is
given in Appendix~\ref{app:hobson}.

In this paper, we describe a fast gridded method for the $uv$-plane
analysis of large interferometric data sets.  The basis of this approach
is to grid the visibilities and perform maximum
likelihood estimation of the power spectrum on this compressed data.
Our use of gridded estimators is significantly different from what has been
done previously.  In addition to power spectrum extraction, this procedure
has the ability to form optimally filtered images from the gridded estimators,
and may be of use in interferometric observations of radio sources in general.

We have applied our method to the analysis of data from the Cosmic Background
Imager (CBI).  The CBI is a planar interferometer array of
13 individual 90-cm  Cassegrain antennas on a 6-m pointable platform
\citep{pa02}.  It covers the frequency range 26--36~GHz in 10 contiguous 1~GHz
channels, with a thermal noise level of $2\,\mu$K in 6 hours, 
and a maximum resolution of $4'$ limited by the longest baselines.
The CBI baselines probe $\ell$ in the range 500--3900.
The 90-cm antenna diameters were chosen to maximize sensitivity, but their
primary beamwidth of $45\farcm2$ (FWHM) at 31~GHz limits the instantaneous 
field of view, which in turn limits the resolution in $\ell$.
This loss of aperture plane resolution can be overcome by making mosaic
observations, i.e.\ observations in which a number of adjacent pointings 
are combined \citep{ekers,co88,co93,sault}.  In
the CBI observations, mosaicing a field several times larger than the 
primary beam has resulted in an increase in resolution in $\ell$
by more than a factor of 3, sufficient to discern features in the power
spectrum.

The first CBI results were presented in \citet{paper1}, hereafter Paper~I, 
using earlier versions of the software that did not make use
of $uv$-plane gridding, and were far too slow to be used on larger
mosaiced data sets.  It was therefore essential to develop a more
efficient analysis method that would be fast enough to carry out
extensive tests on the CBI mosaic data.  The software package
described below has been used to process the first year of CBI data.
In the companion papers \citep[hereafter Paper~II]{paper2} and 
\citep[hereafter Paper~III]{paper3}, the results from
passing CBI deep-field data and mosaic data respectively through the pipeline
are presented.  This paper is Paper~IV in the series.
The output from this pipeline is then used to derive
constraints on cosmology \citep[hereafter Paper~V]{paper5}.  Finally, 
analysis of the excess of power at high-$\ell$ seen in results shown in
Paper~II in the context of the Sunyaev-Zeldovich effect
is carried out, again using the method presented here, in 
\citep[hereafter Paper~VI]{paper6}.

An introduction to the properties of the CMB power spectrum, the 
response of an interferometer to the incoming radiation, and the
computation of the primary beam are given in sections \S\ref{sec:pscmb},
\S\ref{sec:basic}, and \S\ref{sec:gauss} respectively.
The gridding process is presented in \S\ref{sec:gridmethod}, followed
by a description of the likelihood function and construction
of the various covariance matrices in \S\ref{sec:liklhd}.  Details on
the maximum likelihood solution and the calculation of window functions and
component bandpowers is given in \S\ref{sec:relax}, while \S\ref{sec:image}
presents our method for making optimally filtered images from the gridded
estimators.  Finally, a description of the CBI implementation of this method
and the performance of the pipeline, including demonstrations using simulated
CBI datasets, is given in \S\ref{sec:estliklhd}, followed by a summary and
conclusions in \S\ref{sec:conclude}.

\section{The CMB Power Spectrum} \label{sec:pscmb}

At small angles, the curvature of the sky is negligible and we can approximate
the spherical harmonic transform of the the temperature field $T(\bvec{x})$
in direction $\bvec{x}$ as its Fourier transform 
$\tilde{T}(\bvec{u})$ \citep{be87}, where $\bvec{u}$ is the
conjugate variable to $\bvec{x}$.  We adopt the Fourier convention
\begin{equation} \label{eq:fourier}
   \tilde{F}(\bvec{u}) = \int d^2\bvec{x}\, F(\bvec{x})\,
   e^{-2\pi i \sbvec{u}\cdot\sbvec{x}} \\
   \hspace{1cm} \Leftrightarrow \hspace{1cm}
   F({\bf x}) = \int d^2\bvec{u}\, \tilde{F}(\bvec{u})\,
   e^{2\pi i \sbvec{u}\cdot\sbvec{x}}
\end{equation}
of \citet{bracewell}, \citet{tms}, and \citet{synim2}.  
In terms of the multipoles $\ell$,
\begin{equation} \label{eq:celldef}
   \left\langle \tilde{T}(\bvec{u})^2 \right\rangle  \approx C_\ell
   \hspace{2cm} \ell + 1/2 \approx 2\pi |\bvec{u}|
\end{equation}
which we simplify to $\ell = 2\pi |\bvec{u}|$ 
for the $\ell>100$ of interest in this paper.  For the low levels
anisotropy seen in the CMB on these scales, it is convenient to give
$T$ in units of $\mu$K and thus $C_\ell$ will be in units of $\mu{\rm K}^2$.

Because the CMB is assumed to be a statistically homogeneous Gaussian random
field, the components of its Fourier transform are independent Gaussian
deviates.
\begin{equation} \label{eq:cofu}
  \left\langle \tilde{T}(\bvec{u})\,\tilde{T}^\ast(\bvec{u}') \right\rangle =
    C(|\bvec{u}|)\, \delta^2( \bvec{u} - \bvec{u}' ) 
\end{equation}
where $C(|\bvec{u}|) = C_{2\pi |\sbvec{u}|}$.  
Because $T(\bvec{x})$ is real, its transform must be Hermitian, with
$\tilde{T}(\bvec{u}) = \tilde{T}^\ast(-\bvec{u})$, and therefore
\begin{equation} \label{eq:cofualt}
  \left\langle \tilde{T}(\bvec{u})\,\tilde{T}(\bvec{u}') \right\rangle =
  \left\langle \tilde{T}(\bvec{u})\,\tilde{T}^\ast(-\bvec{u}') \right\rangle =
  C(|\bvec{u}|)\, \delta^2( \bvec{u} + \bvec{u}' ).
\end{equation}
Note that it is common to write the CMB power spectrum $C_\ell$ in a form 
\begin{equation} \label{eq:cvflat}
   {\cal C}_\ell = { \ell(\ell+1) \over 2\pi }\,C_\ell 
     \approx{ \ell^2 \over 2\pi }\,C_\ell
   \hspace{1cm} \Leftrightarrow \hspace{1cm}
   {\cal C}(|\bvec{u}|) \approx 2\pi\,|\bvec{u}|^2\,C(|\bvec{u}|) 
\end{equation}
\citep{wcdh99,bjk98,bjk00}.
Constant ${\cal C}$ corresponds to equal power in equal intervals of
$\log\ell$.

Although the power spectrum $C_\ell$ is defined in units of 
brightness temperature, the interferometer measurements carry the units of flux
density $S_\nu$ (Janskys, 1~Jy = $10^{-26}$ W m$^{-2}$ Hz$^{-1}$). In
particular, the intensity field on the sky $I_\nu(\bvec{x})$ 
has units of specific intensity (W m$^{-2}$ Hz$^{-1}$ sr$^{-1}$ or Jy/sr), and
thus to convert between $I_\nu$ and $T$ we use 
$I_\nu(\bvec{x}) = f_{\rm T}(\nu)\,T(\bvec{x})$ with the Planck factor 
\begin{equation} \label{eq:fplanck}
   f_{\rm T}(\nu) = 2\nu^2 k_{\rm B} g(\nu, T_0) / c^2 \hspace{2cm}
   g(\nu, T_0) = { x^2 e^x \over (e^x - 1)^2 } \hspace{2cm}
   x = h \nu / k_{\rm B} T_0
\end{equation}
where $g$ corrects for the blackbody spectrum.  Note that
We have treated the temperature $T$ as small fluctuations about the
mean CMB temperature $T_0=2.725$~K \citep{mather}, and thus the $g$
appropriate to $T_0$ is used with $g\approx0.98$ at $\nu=31$~GHz.

We are not restricted to modeling the CMB.  For
example, we might wish to determine the power spectrum of fluctuations in a
diffuse galactic component such as synchrotron emission or thermal dust
emission. In this case, one might wish to express $I$ in Jy/sr, but take out a
power-law spectral shape 
\begin{equation} \label{eq:alpha}
  I_\nu = f_0(\nu)\,I_0
  \hspace{2cm}
  f_0(\nu) = \left( { \nu \over \nu_0 } \right)^{\alpha} 
\end{equation}
where $\alpha$ is the spectral index, and $f_0(\nu)$ is the conversion
factor that normalizes to the intensity $I_0$ at the 
fiducial frequency $\nu_0$.  Note that this normalization is particularly
useful for fitting out centimeter-wave foreground emission, which tends to
have a power-law spectral index in the range $-1 < \alpha < 1$ that is
significantly different from that for the thermal CMB ($\alpha \approx 2$).
In addition, foregrounds will also tend to have a power spectrum shape
different from that of CMB, which must be included in the analysis (see
\S\ref{sec:foreground} below).

\section{Response of the Interferometer} \label{sec:basic}

A visibility $V_k$ formed from the correlation of an interferometer baseline
between two antennas with projected separation (in the plane perpendicular
to the source direction) $\bvec{b}$ meters observed at wavelength $\lambda$
meters measures (in the absence of noise) the Fourier transform of the sky
intensity modulated by the response of the antennas \citep{tms}
\begin{equation} \label{eq:point1}
  V(\bvec{u}) = \int d^2\bvec{x}\, {\cal A}(\bvec{x})\,I(\bvec{x})\,
   e^{-2\pi i \sbvec{u}\cdot\sbvec{x}}
   \hspace{2cm} \bvec{u} = \bvec{b}/\lambda
\end{equation}
where ${\cal A}(\bvec{x})$ is the primary beam, and $\bvec{u}=(u,v)$ is
the conjugate variable to $\bvec{x}$.  For angular coordinates
$\bvec{x}$ in radians, $\bvec{u}$ has the dimensions
of the baseline or aperture in units of the wavelength. 
The Fourier domain is also referred to as the {\it uv}-plane or aperture
plane in interferometry for this reason.

We define the direction cosines
\begin{eqnarray} \label{eq:xi}
  \bvec{x}_k = (\Delta x_k,\Delta y_k) \qquad 
  \Delta x_k & = & \cos\delta_k \, \sin(\alpha_k - \alpha_0) \nonumber \\
  \Delta y_k & = & \sin\delta_k\,\cos\delta_0 - 
       \cos\delta_k\,\sin\delta_0\,\cos(\alpha_k - \alpha_0)
\end{eqnarray}
between the point at right ascension and declination
$\alpha_k,\delta_k$ and the center of the mosaic $\alpha_0,\delta_0$.
For the CBI, data are taken keeping the 
phase center fixed on the pointing center $\bvec{x}_k$ by 
shifting the phase with the beam and rotating the platform to
maintain constant parallactic angle during a scan, so that the
response to a point source at the center of the field
$I(\bvec{x}) = \delta^2(\bvec{x}-\bvec{x}_k)$ is constant, and thus
\begin{equation}
  {\cal A}(\bvec{x}) = A_{k}(\bvec{x}-\bvec{x}_k)\,
  e^{2\pi i \sbvec{u}_k\cdot\sbvec{x}_k}
\end{equation}
in equation (\ref{eq:point1}), where $A_{k}$ is the normalized primary beam 
response at the observing frequency of visibility $k$.  Then,
by application of the Fourier shift theorem, it is easy to show that
\begin{eqnarray} 
  V_k & = & \int d^2\bvec{x}\, A_{k}(\bvec{x}-\bvec{x}_k)\,
  I_{\nu_k}(\bvec{x})\,e^{-2\pi i \sbvec{u}_k\cdot(\sbvec{x}-\sbvec{x}_k)}
  + e_k \nonumber \\
  & = & \int d^2\bvec{v}\, \tilde{A}_k(\bvec{u}_k-\bvec{v})\,
  \tilde{I}_{\nu_k}(\bvec{v})\,e^{2\pi i \sbvec{v}\cdot\sbvec{x}_k} 
  + e_k
\label{eq:visi}
\end{eqnarray}
where $\tilde{A}_{k}$ is the Fourier transform of the primary beam ${A}_{k}$,
and $I_\nu(\bvec{x})$ is the sky brightness field (expressed
in units such as Jy/sr) with transform $\tilde{I}_\nu(\bvec{v})$.
The instrumental noise on the complex visibility measurement is represented
by $e_k$.

The {\it uv}-plane resolution of an interferometer in a single
pointing is thus limited by the convolution with $\tilde{A}$.  
However, these sub-aperture spatial frequencies can be recovered by
using the phase gradient in the exponential 
$\exp\big[ 2\pi i \bvec{v}\cdot\bvec{x}_k \big]$ from a raster
of mosaic pointings $\{\bvec{x}_k\}$, provided that the spacing of the
pointings is sufficiently small to avoid aliasing as discussed in
Appendix~\ref{app:estim}.

To aid us later on, we introduce a convolution kernel
\begin{equation} \label{eq:pk}
  P_k(\bvec{v}) = f_k\,\tilde{A}_k(\bvec{u}_k-\bvec{v})\,
  e^{2\pi i \sbvec{v}\cdot\sbvec{x}_k}
\end{equation}
and thus
\begin{equation} \label{eq:vispk}
  V_{k} = \int d^2\bvec{v}\, P_k(\bvec{v})\,\tilde{T}(\bvec{v}) + e_k
\end{equation}
where $f_k=f_{\rm T}(\nu_k)$ is the Planck conversion factor for the CMB given
in equation (\ref{eq:fplanck}).
It is easiest to write these in operator notation, with
\begin{equation} \label{eq:visop}
  \bvec{V} = {\bf P}\,\tilde{\bvec{T}} + \bvec{e}
\end{equation}
where ${\bf V}$ and ${\bf e}$ are the visibility and noise vectors
respectively, ${\bf P}$ is our kernel,
and $\tilde{\bvec{T}}$ is the transform of the temperature field.  In this
representation $\tilde{\bvec{T}}$ can be thought of as a vector of
cells in {\it uv} space.

\section{The Primary Beam} \label{sec:gauss}

In order to determine the response of the array to the radiation field,
we need to know the primary beam $A(\bvec{x})$ of the antenna elements
and its Fourier transform $\tilde{A}(\bvec{u})$.  In general, for
each frequency channel, each baseline has a primary beam which is the Fourier
transform of the cross-correlation of the voltage illumination functions
across the aperture of each antenna --- see \citet{tms} for a detailed
treatment of the interferometer response.  For a real and symmetric primary
beam that is identical between antennas, then the transforms are symmetric and
real we can ignore the differences between cross-correlation and convolution
and write 
\begin{equation}
   \tilde{A}(\bvec{u}) = \hat{g} \star \hat{g}
   \Leftrightarrow 
   A(\bvec{x}) = \big| \tilde{g}^2 \big|
\end{equation}
for the voltage illumination function $\hat{g}(r,\nu)$ across
the radius of the aperture $r=|\bvec{r}|$ at frequency $\nu$, where 
$\tilde{g}$ is the Fourier transform of $\hat{g}$.  The CBI beams have
been measured and are nearly identical and symmetric, and thus we will
use a single mean primary beam and its transform for the array.  For
a heterogeneous array, the individual beams can be used with some added
complication.

For most antennas, such as those used in the CBI, the primary beam
width scales linearly with observing wavelength, and thus $\hat{g}(r)$ is
approximately constant with wavelength.  Then, we can 
define $G(r)$ as the normalized aperture autocorrelation function,
and write
\begin{equation} \label{eq:grade}
   \tilde{A}_k(\bvec{u}) = {1\over A_0}\,G( |\bvec{u}|\,\lambda_k )
\end{equation}
for a channel centered at wavelength $\lambda_k$, with 
\begin{equation} \label{eq:anorm}
  A_0 = \int d^2\bvec{u}\,G( |\bvec{u}|\,\lambda_k )
      = {2\pi\over\lambda_k^2}\int_0^{\infty} r\,dr\,G( r )
\end{equation}
normalizes the response to give unity gain on-sky at the beam center
($A(0)\equiv 1$).  If $g(r)=\hat{g}(r)/g(0)$, then $G(r)=g\star g$.

The two-dimensional primary beam response, $A(\bvec{x})$, 
is determined by means of measurements of a bright radio source over a
two-dimensional grid of offset pointings centered on the source.  
The central lobe of $A(\bvec{x})$ for the CBI is well approximated by
a circular Gaussian, which is characterized by its dispersion $\sigma_x$, 
which is related to the full width at half-maximum (FWHM) $a_x$ by
$\sigma_x = a_x/\sqrt{8\,\ln2}$.  The Fourier transform
of an infinite circular Gaussian is given by
\begin{equation} \label{eq:gbeam}
  A(\bvec{x}) = e^{ - { x^2 \over 2\sigma_x^2}} \Leftrightarrow
  \tilde{A}(\bvec{u}) = {1 \over 2 \pi \sigma_u^2}\,
  e^{ - { |\sbvec{u}|^2 \over 2\sigma_u^2}} \hspace{2cm}
  \sigma_u = { 1 \over 2 \pi \sigma_x }
\end{equation}
where $\sigma_u$ is the Gaussian dispersion in Fourier space.
The function $G(r)$ is therefore
\begin{equation} \label{eq:fbeam}
   G(r) = e^{ -{r^2 \over 2r_g^2} }
\end{equation}
for Gaussian radius $r_g=\lambda\,\sigma_u$.
For the CBI the measured primary beam (see Paper~III) has
$a_x=45\farcm2\times(31{\rm 
GHz}/\nu)$, so $\sigma_u = 28.50$ at 31~GHz ($\lambda=0.967$~cm),
which corresponds to $r_g=27.56$~cm.  

For the CBI software pipeline, instead of using a Gaussian approximation
to $G(r)$ we have chosen to model the antenna illumination $g(r)$ as a
Gaussian truncated at both the dish edge and the secondary blockage radius 
\begin{equation} \label{eq:tillum}
  g(r) = \left\{ \begin{array}{ll}
    0 & |r| \leq r_{\rm inner} \\
    e^{ - { \left({r \over s}\right)^2 } } & r_{\rm inner} < |r| < D/2 \\
    0 & |r| \geq D/2 \end{array} \right.
\end{equation}
where for the CBI antennas $r_{\rm inner}=7.75$~cm.  Note that if $g(r)$ and 
$G(r)$ were infinite circular Gaussians, then $s=r_g$.  A best-fit taper
parameter $s$ is obtained using the measured primary beam $A$, giving
$s=30.753$~cm or an edge taper of 0.118 ($-18.6$~db of power) at the dish
edge.  We then numerically tabulate the autocorrelation $G(r)$
assuming $s=30.753$~cm which is then interpolated in the code
when $\tilde{A}$ is required.  This model is a better fit to the observed
beam than a pure Gaussian beam (see Figure~1 in Paper~III for a plot
of this model).

The resolution in $uv$ or $\ell$ space is set by the width of 
$\tilde{A}_k(\bvec{u})$.  For a Gaussian approximation to the beam, the
dispersion in multipole $\ell$ is $\sigma_\ell = 2\pi\sigma_u = 1/\sigma_x$,
and the FWHM is $a_\ell = 8\,\ln 2/a_x$.  For $a_x = 45\farcm2$ at 31~GHz we
have FWHM $a_\ell=422$ ($\sigma_\ell=179$).  Given that features
are expected in the power spectrum of widths significantly less than
this, it is highly desirable to reduce the effective resolution width of
the CBI by at least a factor three using mosaicing.

\section{Gridded Estimators}\label{sec:gridmethod}

The principal problem in using likelihood (see \S\ref{sec:liklhd})
to determine confidence limits
on the power spectrum for CBI data is the large number of visibilities
compounded by the large number of mosaic pointings (typically $7\times6$
or larger).  Even a modest reduction in the number of matrix elements
passed to the likelihood calculation will greatly aid the computation.  This
suggests that we grid the visibilities before
computing the likelihood function.  For an effective resolution
in the aperture plane determined by the primary beam and mosaic size,
there is little use in sampling below this smearing scale, 
and we can define an optimum gridding scheme
which minimizes the quantity of data and information loss (the gridding
is a form of compression).

We implement this by defining {\it estimators} $\Delta(\bvec{u})$ for the
true complex brightness transform which are linear combinations of the
measured visibilities. These estimators bin together data from the
different frequency bands and mosaic pointings.  Thus, a direct sum of
visibilities taken at the same ${\bf u}$ but over the whole mosaic ${\bf x}$
will result in an estimator that has a higher effective resolution in the
{\it uv} plane.  The result is that we can speed up the likelihood 
estimation at the cost of complicating the covariance matrix.
In general, this matrix can be computed relatively quickly as it is a
$N^2$ process, and thus this is a worthwhile trade-off versus the $N^3$
cost of calculating the likelihood.  The estimators derived in 
Appendix~\ref{app:estim} are not orthogonal combinations of the original
visibilities, and thus some information loss is expected.  However, the
tests performed in \S\ref{sec:test} show that these estimators are unbiased,
and comparisons to results obtained using the visibilities directly show
that there is no noticeable loss in sensitivity.  Thus, our gridding can
be considered to be an efficient form of (lossy) compression using the beam as
a signal template.

In Appendix~\ref{app:estim}, we argue that a $\Delta_i$ formed by a linear
combination of visibilities will give a estimate of the weighted average of
$\tilde{I}$ or $\tilde{T}$ around {\it uv} locus $\bvec{u}_i$.  
We have from equation (\ref{eq:moskern})
\begin{equation} \label{eq:linop}
   \bvec{\Delta} = {\bf Q}\,\bvec{V} + \overline{\bf Q}\,\bvec{V}^\ast 
\end{equation}
where the kernel ${\bf Q}$ is defined in equation (\ref{eq:moskern}) and
the kernel for the conjugate visibilities
$\overline{\bf Q}$ is defined in equation (\ref{eq:moskernalt}).  In 
particular,
\begin{eqnarray}
   Q_{ik} & = & { \omega_k \over z_i }\,
      \tilde{A}^\ast_k(\bvec{u}_k-\bvec{u}_i) 
      e^{-2\pi i \sbvec{u}_i\cdot\sbvec{x}_k} \nonumber \\
   \overline{Q}_{ik} & = & { \omega_k \over z_i }\,
      \tilde{A}^\ast_k(-\bvec{u}_k-\bvec{u}_i) 
      e^{-2\pi i \sbvec{u}_i\cdot\sbvec{x}_k}
      \label{eq:qik}
\end{eqnarray}
where $z_i$ the normalization factor given 
in equation (\ref{eq:zk1}), and $\omega_k=\epsilon_k^{-2}$ is the 
visibility weight given in equation (\ref{eq:viswt}).	

By operating with the gridding kernel on equation (\ref{eq:visop}), we get 
\begin{equation}
   \bvec{\Delta} = {\bf R}\,\tilde{\bvec{T}} + \bvec{n}
   \hspace{2cm}
   {\bf R} = {\bf Q}\,{\bf P} + \overline{\bf Q}\,\overline{\bf P}
   \hspace{2cm}
   \bvec{n} = {\bf Q}\,\bvec{e} + \overline{\bf Q}\,\bvec{e}^\ast 
   \label{eq:rop}
\end{equation}
where we define ${\bf R}$ as the convolution kernel that operates on the
transform of the intensity (the gridded version of ${\bf P}$), and
$\bvec{n}$ is the gridded noise.
The conjugate to ${\bf P}$ defined in equation (\ref{eq:pk}) is given by
\begin{equation} \label{eq:altpfull}
   \overline{P}_{k}(\bvec{v}) = f_k\,\tilde{A}_k(-\bvec{u}_k-\bvec{v})\,
      e^{2\pi i \sbvec{v}\cdot\sbvec{x}_k},
\end{equation}
which gathers the conjugate visibilities under the transformation 
$\bvec{u}_k \rightarrow -\bvec{u}_k$.

Although it is not necessary to do so, it is convenient to construct
the $\Delta_i$ on a regular lattice in $\bvec{u}_i$ with a spacing
$d_u$.  Thus the grid ``cells'' represented by the $\Delta_i$ represent an
interpolation using ${\bf Q}$ of the visibilities onto the {\it uv}-plane.
This will be useful when using the estimators to form filtered images
(\S\ref{sec:image}).

If it is desired that the visibilities be used directly, for example when the
datasets are small, then the ungridded case can be recovered by setting
$Q_{ik}=\delta_{ik}$ and $\overline{Q}_{ik}=0$, giving $\bvec{\Delta} =
\bvec{V}$ and ${\bf R} = {\bf P}$, with no loss of generality in the
derivations.  

\section{The Likelihood Function} \label{sec:liklhd}

To carry out the power spectrum estimation, we form the likelihood of
the data given covariance matrices for the signal, noise, and foregrounds.
Since the estimators are linear combinations of the visibilities, which
we assume are made up of Gaussian noise and Gaussian signal components,
we can use a multivariate Gaussian probability distribution to describe
the estimators also.  Because ${\bf \Delta}$ is complex, it is easier to deal
with the real and imaginary parts by packing them together in a double-length
real vector
\begin{equation} \label{eq:xvec}
  \bvec{d} =  \left( \begin{array}{c} 
     {\rm Re}\,\bvec{\Delta} \\
     {\rm Im}\,\bvec{\Delta} \end{array} \right)
\end{equation}
written here as a column-vector of length $2N_{\rm est}$.

The log-likelihood function for a real multivariate Gaussian
probability distribution is 
\begin{equation} \label{eq:likx}
  \ln L(\bvec{x}|\bvec{q}) = -N_{\rm est}\,\ln 2\pi - 
  {1\over2}\,\ln ( \det {\bf C} ) - 
  {1\over2}\,\bvec{d}^{\rm t}\,{\bf C}^{-1}\,\bvec{d}
\end{equation}
where $\bvec{d}^{\rm t}$ is the transpose of $\bvec{d}$, and 
\begin{equation} \label{eq:creim}
  {\bf C} =
  \left( \begin{array}{cc} 
    \langle{\rm Re}\,\bvec{\Delta}\,{\rm Re}\,\bvec{\Delta}^{\rm t} \rangle & 
    \langle{\rm Re}\,\bvec{\Delta}\,{\rm Im}\,\bvec{\Delta}^{\rm t} \rangle \\
    \langle{\rm Im}\,\bvec{\Delta}\,{\rm Re}\,\bvec{\Delta}^{\rm t} \rangle &
    \langle{\rm Im}\,\bvec{\Delta}\,{\rm Im}\,\bvec{\Delta}^{\rm t} \rangle 
  \end{array} \right)
\end{equation}
is a block matrix of the real and imaginary covariances.
The vector $\bvec{q}$ represents the parameters of the model or theory
against which the data is being measured (see below).  These parameters
are contained in ${\bf C}$.

In terms of $\bvec{\Delta}$ and $\bvec{\Delta}^\ast$, we can write
\begin{equation} \label{eq:vreim}
 {\rm Re}\,\bvec{\Delta} = 
    {1\over2}\,\left( \bvec{\Delta} + \bvec{\Delta}^\ast \right)
 \hspace{2cm}
 {\rm Im}\,\bvec{\Delta} = 
    {1\over2i}\,\left( \bvec{\Delta} - \bvec{\Delta}^\ast \right)
\end{equation}
and therefore
\begin{eqnarray} 
\langle {\rm Re}\,\bvec{\Delta}\,{\rm Re}\,\bvec{\Delta}^{\rm t} \rangle & = & 
  \phantom{-}{1\over2}\,{\rm Re}\,\left[ 
  \langle \bvec{\Delta}\,\bvec{\Delta}^\dag \rangle + 
  \langle \bvec{\Delta}\,\bvec{\Delta}^{\rm t} \rangle \right] \nonumber \\
\langle {\rm Im}\,\bvec{\Delta}\,{\rm Im}\,\bvec{\Delta}^{\rm t} \rangle & = &
  \phantom{-}{1\over2}\,{\rm Re}\,\left[ 
  \langle \bvec{\Delta}\,\bvec{\Delta}^\dag \rangle - 
  \langle \bvec{\Delta}\,\bvec{\Delta}^{\rm t} \rangle \right] \nonumber \\
\langle {\rm Re}\,\bvec{\Delta}\,{\rm Im}\,\bvec{\Delta}^{\rm t} \rangle & = &
 -{1\over2}\,{\rm Im}\,\left[ \langle {\bf \Delta}\,{\bf \Delta}^\dag \rangle
  - \langle \bvec{\Delta}\,\bvec{\Delta}^{\rm t} \rangle \right] \nonumber \\
\langle {\rm Im}\,\bvec{\Delta}\,{\rm Re}\,\bvec{\Delta}^{\rm t} \rangle & = &
  \phantom{-}{1\over2}\,{\rm Im}\,\left[ 
  \langle \bvec{\Delta}\,\bvec{\Delta}^\dag \rangle 
  + \langle \bvec{\Delta}\,\bvec{\Delta}^{\rm t} \rangle \right] 
  \label{eq:cimre}
\end{eqnarray}
where $\bvec{\Delta}^\dag=(\bvec{\Delta}^\ast)^{\rm t}$ is the Hermitian 
transpose of $\bvec{\Delta}$ (a row-vector containing the complex conjugate of
a column-vector), and $\bvec{\Delta}\,\bvec{\Delta}^\dag$ is the tensor or
outer product of $\bvec{\Delta}$ and $\bvec{\Delta}^\dag$, which is a matrix
with elements $\Delta_i\,\Delta_j^\ast$.

\begin{figure}
\includegraphics[width=6.5in]{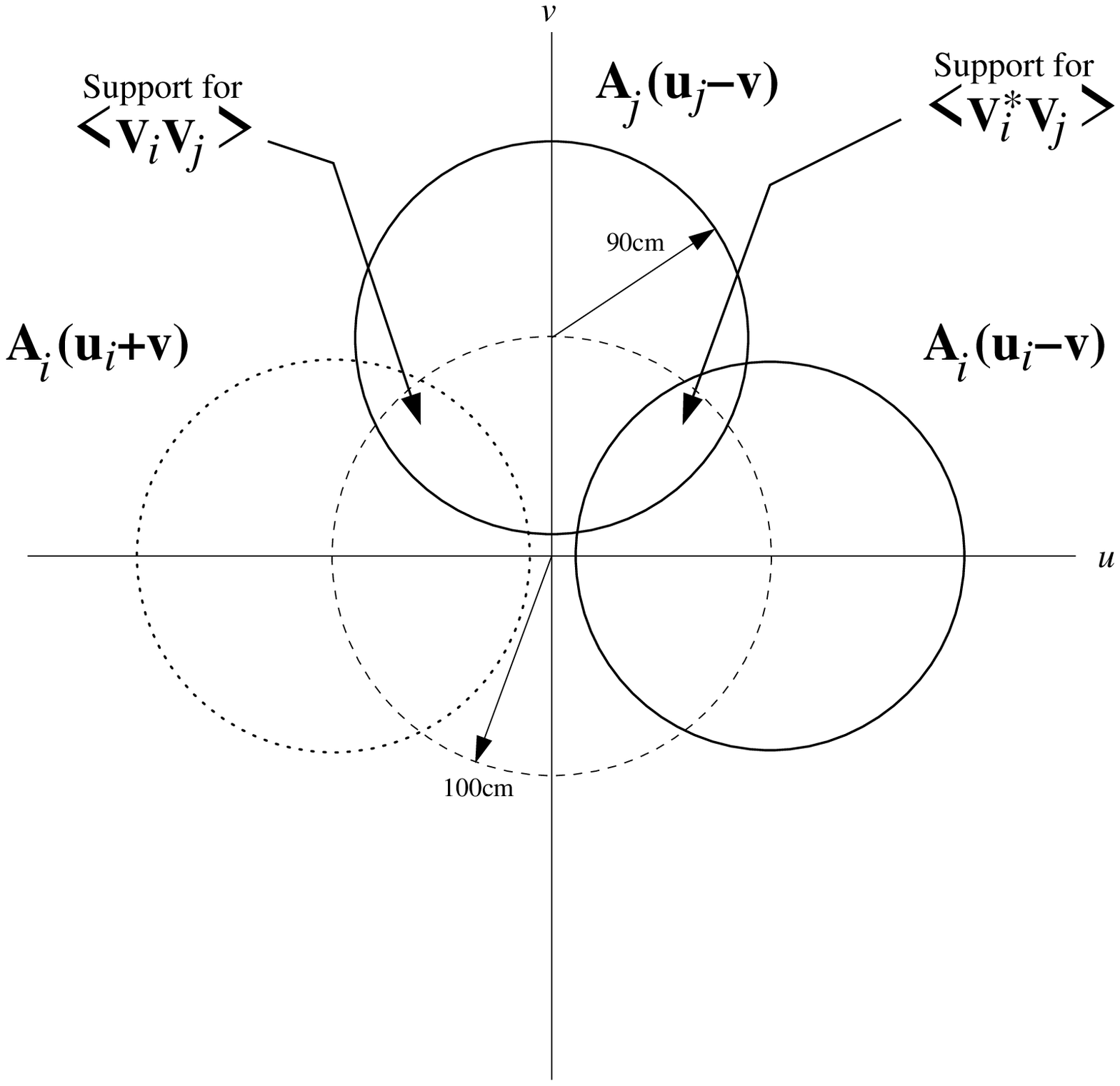}
\caption{Graphical representation of the regions of support in the
aperture plane for the correlation between visibilities on short
baselines $B<\sqrt{2}D$.  Note that both $V_i$ and its conjugate
$V^\ast_i=V(-{\bf u}_i)$ have overlapping support for visibility
$V_j$, and this must be taken into account in computing the covariance
matrix element.\label{fig:support}}
\end{figure}

It is important to include the covariances of
$\bvec{\Delta}\,\bvec{\Delta}^{\rm t}$ 
as well as those for $\bvec{\Delta}\,\bvec{\Delta}^\dag$.
Normally, only one of a given visibility $V_k$ or its conjugate $V^\ast_k$ will
correlate with $V_{k'}$. However, for short baselines $b<\sqrt{2}\,D$ (less
than 127.3~cm for the 90~cm CBI dishes), there may be overlap between the
support for a given visibility and both another visibility and its conjugate,
as shown in Figure~\ref{fig:support}, and thus both may be nonzero.  Note that
the correlation between distant conjugate pairs is small,
since the overlap occurs far out in the antenna response $\tilde{A}$, although
it is non-negligible on the shortest CBI baselines where the overlap occurs at
the $0.57 D$ point (illustrated in Figure~\ref{fig:support}) for perpendicular
1-meter baselines with the beam response $\sim 30\%$.  Outside the baseline
range $b>\sqrt{2}\,D$ one of $\langle V^\ast_{k}\, V_{k'} \rangle$ or $\langle
V_{k}\, V_{k'} \rangle$ will be zero.

The covariance matrix ${\bf C}$ can be split into a sum of independent
contributions from instrumental noise ${\bf C}^{\rm N}$, the CMB signal ${\bf
C}^{\rm S}$, and foreground signals ${\bf C}^{\rm src}$ and ${\bf C}^{\rm
res}$.  We further split ${\bf C}^{\rm S}$ into a sum of terms 
${\bf C}^{\rm S}_B$ from separate $\ell$ bands of the power spectrum, 
\begin{equation} \label{eq:csnbandv}
   {\bf C} = {\bf C}^{\rm N} + \sum_B q_B\,{\bf C}^{\rm S}_B
   + q_{\rm src}\,{\bf C}^{\rm src} + q_{\rm res}\,{\bf C}^{\rm res}.
\end{equation}
The factors $\{ q_B, B =1,\ldots,N_B \}$ are the ``bandpowers''
\citep{bjk98} for bins with centers at $\ell = \ell_B$, and are
the model parameters to be determined by maximizing the likelihood.
The factor $q_{\rm res}$ is amplitude of the covariance due to a residual
Gaussian foreground, and $q_{\rm src}$ is the amplitude of the covariance
contributed by known point sources; there may more than one of each
of these types of foreground covariance matrices.  The $q_{\rm src}$ and
$q_{\rm res}$ can be treated as adjustable parameters to determined by maximum
likelihood, or they can be held fixed at {\it a priori} values, in which case
${\bf C}^{\rm src}$ and ${\bf C}^{\rm res}$ are constraint matrices with their
corresponding terms in equation (\ref{eq:csnbandv}) behaving
like additional noise terms.

In the following sections we consider each of the terms ${\bf C}^{\rm N}$,
${\bf C}^{\rm S}_B$, ${\bf C}^{\rm src}$, and ${\bf C}^{\rm res}$ in turn.
If we write
\begin{equation} \label{eq:mmbar}
  {\bf M}=\langle \bvec{\Delta}\,\bvec{\Delta}^\dag \rangle
  \hspace{2cm}
  \overline{\bf M}=\langle \bvec{\Delta}\,\bvec{\Delta}^{\rm t} \rangle
\end{equation}
then in each case we calculate the contributions  
to the covariance matrix for the real and imaginary parts of the
estimators using equations (\ref{eq:creim}) and (\ref{eq:cimre})
\begin{equation} \label{eq:csbreimest}
  {\bf C} = \left( \begin{array}{cc}
  \phantom{-}{1\over2}\,{\rm Re}\,[ {\bf M} + \overline{\bf M} ] & 
            -{1\over2}\,{\rm Im}\,[ {\bf M} - \overline{\bf M} ] \\
  \phantom{-}{1\over2}\,{\rm Im}\,[ {\bf M} + \overline{\bf M} ] &
  \phantom{-}{1\over2}\,{\rm Re}\,[ {\bf M} - \overline{\bf M} ] 
\end{array} \right)
\end{equation}
with the individual covariance matrices given by insertion of the
appropriate contribution to ${\bf M}$ and $\overline{\bf M}$ for that
component, e.g.\ ${\bf M}^{\rm S}_B$ and $\overline{\bf M}^{\rm S}_B$ to
compute the block elements of ${\bf C}^{\rm S}_B$.

\subsection{The Noise Covariance Matrix} \label{sec:noisecor}

The instrumental noise correlations are assumed to be Gaussian and independent
between different baselines, and frequency channels.  For the CBI, tests
have been carried out on the data which show this to be true to a high
level of accuracy.  In this case, the noise contribution
to the real and imaginary parts of the visibilities are independent zero-mean
normal deviates with
\begin{equation}
  \langle {\rm Re}\,e_{k}\,{\rm Re}\,e_{k'} \rangle = 
  \langle {\rm Im}\,e_{k}\,{\rm Im}\,e_{k'} \rangle = 
  \epsilon_k^2\,\delta_{kk'}
  \hspace{2cm}
  \langle {\rm Re}\,e_{k}\,{\rm Im}\,e_{k'} \rangle = 0
\end{equation}
and thus we can write 
\begin{equation}
  \langle \bvec{e}\,\bvec{e}^\dag \rangle = {\bf E}
  \hspace{2cm} \langle \bvec{e}\,\bvec{e}^{\rm t} \rangle = 0
  \label{eq:ndef}
\end{equation}
for real noise matrix ${\bf E}$, where $E_{kk'} = 2\epsilon^2_k\,\delta_{kk'}$.

It can be shown that the noise contributions $\bvec{n}$ to the estimators
defined in equation (\ref{eq:rop}) have the contributions
to the covariance elements ${\bf M}$ and $\overline{\bf M}$ defined
in equation (\ref{eq:mmbar}) given by
\begin{equation} \label{eq:eeexmat}
  \begin{array}{ccccc}
    {\bf M}^{\rm N} & = & \langle \bvec{n}\,\bvec{n}^\dag \rangle 
      & = & {\bf Q}\,{\bf E}\,{\bf Q}^\dag + 
            \overline{\bf Q}\,{\bf E}\,\overline{\bf Q}^\dag \\
    \overline{\bf M}^{\rm N} & = & \langle \bvec{n}\,\bvec{n}^{\rm t} \rangle
      & = & {\bf Q}\,{\bf E}\,\overline{\bf Q}^{\rm t} +
      \overline{\bf Q}{\bf E}\,{\bf Q}^{\rm t}
  \end{array}
\end{equation}
using the covariances of $\bvec{e}$ given in equation (\ref{eq:ndef}).
This is assembled into the covariance matrix ${\bf C}^{\rm N}$ 
using equation (\ref{eq:csbreimest}).  In general, the gridding kernel
${\bf Q}$ will map a given visibility to more than one estimator, and
thus ${\bf C}^{\rm N}$ will have non-zero off-diagonal elements.  
Furthermore, if there are noise
correlations between baselines or channels, then the structure of 
${\bf C}^{\rm N}$ will be even more complicated.

\subsection{The CMB Signal Covariance Matrix} \label{sec:viscor}

The CMB contribution to the visibility covariance matrix is
given by the covariance of the ${\bf R}\,\tilde{\bvec{T}}$ term in
equation (\ref{eq:rop})
\begin{equation} \label{eq:mdef}
  {\bf M}^{\rm S} = {\bf R}\,
    \langle \tilde{\bvec{T}}\,\tilde{\bvec{T}}^\dag \rangle
    \,{\bf R}^\dag 
  \hspace{2cm}
  \overline{\bf M}^{\rm S} = {\bf R}\,
    \langle \tilde{\bvec{T}}\,\tilde{\bvec{T}}^{\rm t} \rangle
    \,{\bf R}^{\rm t}
\end{equation}
where $\langle \tilde{\bvec{T}}\,\tilde{\bvec{T}}^\dag \rangle$ 
and $\langle \tilde{\bvec{T}}\,\tilde{\bvec{T}}^{\rm t} \rangle$ are given 
in equations (\ref{eq:cofu}) and (\ref{eq:cofualt})respectively.
Then, the elements of ${\bf M}^{\rm S}$ and $\overline{\bf M}^{\rm S}$ for
estimators $i$ and $j$ are
\begin{eqnarray} 
   M^{\rm S}_{ij} & = & \int d^2\bvec{v}\,C(|\bvec{v}|) 
   \,R_i(\bvec{v})\,R^\ast_{j}(\bvec{v})
   = {2\pi}\,\int d\varpi\,\varpi\,C(\varpi)\,W_{ij}(\varpi)
   \nonumber \\
   \overline{M}^{\rm S}_{ij} & = & 
   \int d^2\bvec{v}\,C(|\bvec{v}|)\,R_i(\bvec{v})\,R_{j}(-\bvec{v})
   = {2\pi}\,\int d\varpi\,\varpi\,C(\varpi)\,\overline{W}_{ij}(\varpi)
   \label{eq:mkkint}
\end{eqnarray}
with
\begin{eqnarray}
   W_{ij}(\varpi) & = & {1\over 2\pi}\,\int_0^{2\pi} d\theta\,  
   R_{i}(\varpi,\theta)\,R^\ast_{j}(\varpi,\theta) \nonumber \\
   \overline{W}_{ij}(\varpi) & = & {1\over 2\pi}\,\int_0^{2\pi} d\theta\,  
   R_{i}(\varpi,\theta)\,R_{j}(\varpi,\theta-\pi)
   \label{eq:wijvarpi}
\end{eqnarray}
where to aid in breaking up the CMB covariance matrices into bands
we write the integrations in terms of polar Fourier coordinates 
$(u,v)\rightarrow(\varpi,\theta)$ ($\varpi=|\bvec{v}|$).

As an illustration, consider the case without gridding.  Then, 
${\bf R} = {\bf P}$, and using equation (\ref{eq:pk}) in 
(\ref{eq:mkkint}) we get
\begin{eqnarray}
   M^{\rm S}_{kk'} & = & f_k \, f_{k'}\,\int d^2\bvec{v}\,C(|\bvec{v}|)\,
   \tilde{A}_k(\bvec{u}_k-\bvec{v}) \, 
   \tilde{A}^\ast_{k'}(\bvec{u}_{k'}-\bvec{v}) 
   e^{2\pi i \sbvec{v}\cdot(\sbvec{x}_{k} - \sbvec{x}_{k'}) }
   \nonumber \\
   \overline{M}^{\rm S}_{kk'} & = & 
   f_k \, f_{k'}\,\int d^2\bvec{v}\,C(|\bvec{v}|)\,
   \tilde{A}_k(\bvec{u}_k-\bvec{v})\,\tilde{A}_{k'}(\bvec{u}_{k'}+\bvec{v}) 
   e^{2\pi i \sbvec{v}\cdot(\sbvec{x}_k - \sbvec{x}_{k'}) }
   \label{eq:mkkintvis} 
\end{eqnarray}
for the covariance matrix element between visibilities $V_k$ and $V_{k'}$.

We furthermore write the radial integral over $\varpi=\ell/2\pi$ as a sum
with respect to ${\cal C}_\ell$ of equation (\ref{eq:cvflat})
\begin{eqnarray}
   M^{\rm S}_{ij} = \sum_\ell {W_{\ell ij} \over \ell}\,{\cal C}_\ell
   & & W_{\ell ij} = W_{ij}(\ell/2\pi)
   \nonumber \\
   \overline{M}^{\rm S}_{ij} = 
   \sum_\ell {\overline{W}_{\ell ij} \over \ell}\,{\cal C}_\ell
   & & \overline{W}_{\ell ij} = \overline{W}_{ij}(\ell/2\pi) 
   \label{eq:wincor2}
\end{eqnarray}
where $W_{\ell ij}$ is the {\it variance window function} 
(e.g.\ \citealt{knox}).  

We define the bandpowers $\{ q_B, B =1,\ldots,N_B \}$ by constructing
${\cal C}_\ell$ piecewise with respect to a fiducial shape 
${\cal C}^{\rm shape}_\ell$
\begin{equation} \label{eq:cbandell}
 {\cal C}_{\ell} = \sum_B q_B\,{\cal C}^{\rm shape}_{\ell}\,\chi_{B\ell}
\end{equation}
where
\begin{equation} \label{eq:chibl}
 \chi_{B\ell} = \left\{ \begin{array}{ll}
   1 & \ell \in B  \\
   0 & \ell \not\in B  \end{array} \right.
\end{equation}
breaks the power spectrum into non-overlapping bands.  The standard choice for
the shape is ${\cal C}^{\rm shape}_\ell=1$ for equal power 
per log-$\ell$ interval, with $q_B$ then giving the bandpowers in units of
$T^2$.  Then, to calculate ${\bf C}^{\rm S}_B$, we construct band
versions of the covariance matrix elements in equation (\ref{eq:mdef})
\begin{equation}
 {\bf M}^{\rm S}_{B} = \sum_\ell {{\bf W}_{\ell} \over \ell}\,
   {\cal C}^{\rm shape}_\ell \,\chi_{B\ell} 
 \hspace{2cm}
 \overline{\bf M}^{\rm S}_{B} = 
   \sum_\ell {\overline{\bf W}_{\ell} \over \ell}\,
   {\cal C}^{\rm shape}_\ell \, \chi_{B\ell} 
 \label{eq:mkkbin}
\end{equation}
where ${\bf M}^{\rm S} = \sum_B q_B\,{\bf M}^{\rm S}_{B}$ and 
$\overline{\bf M}^{\rm S} = \sum_B q_B\,\overline{\bf M}^{\rm S}_{B}$.  These
are then combined following the prescription in equation (\ref{eq:csbreimest})
to assemble the ${\bf C}^{\rm S}_{B}$.

The variance window function $W_{ij}(\bvec{v})$ is the convolution of the
$\tilde{A}_i(\bvec{v})$ and $\tilde{A}_j(\bvec{v})$, and thus its width is
characteristic of the square of the Fourier transforms primary beam, or
FWHM $\Delta\ell\approx a_\ell/\sqrt{2}$.  
Thus we would expect in a single field to be able to achieve a limiting
resolution of $\Delta\ell \approx 300$ for $a_\ell=422$ at 31~GHz.  This
will be increased by the mosaicing by a factor roughly equal to the extent
of the half-power width of the mosaic relative to that of a single field.
In practice, the limiting useful width for the $\ell$ bins for the bandpowers
will be set by the band-band correlations introduced in the maximum likelihood
estimation procedure (see \S\ref{sec:relax} and \S\ref{sec:test} for
further discussion and examples).

\subsection{Known Point-Source Constraint Matrices}\label{sec:src}

Consider a set of $N_c$ point sources at positions $\bvec{x}_c$ with flux
densities $S_c(\nu)$ ($c=1,\ldots,N_c$).  The intensity field at
frequency $\nu$ is then given by
\begin{equation} \label{eq:isrc}
  I_\nu(\bvec{x}) = \sum_c S_c(\nu)\,\delta^2 ( \bvec{x} - \bvec{x}_c )
\end{equation}
which is assumed to be uncorrelated with other intensity components like
the CMB.  The effect $V^{\rm src}_k$ on the visibilities $V_k$ 
(e.g.\ eq.[\ref{eq:visi}])is then given by the sum over sources
\begin{equation} \label{eq:vsrc}
  V^{\rm src}_k = \sum_c V_{ck} \hspace{2cm}
  V_{ck} = S_c(\nu_k)\,{A}_k(\bvec{x}_c-\bvec{x}_k)\,
    e^{-2\pi i \sbvec{u}_k\cdot(\sbvec{x}_c-\sbvec{x}_k)}
\end{equation}
where $V_{ck}$ is the contribution to visibility $k$ of source $c$.  
We assume that the positions of the sources can be determined with negligible
uncertainty through radio surveys, and that the errors are due to
uncertainties in the measurements of the flux densities.
Then, the covariance between the source contributions to
visibilities $k$ and $k'$ is
\begin{eqnarray}
  \langle V^{\rm src}_{k}\,V^{\rm src \ast}_{k'} \rangle 
    & = & \sum_c \sum_{c'} \langle S_c(\nu_k)\,S_{c'}(\nu_{k'}) \rangle\,
    A_{k}(\bvec{x}_c-\bvec{x}_k)\,A^\ast_{k'}(\bvec{x}_{c'}-\bvec{x}_{k'})\,
  \nonumber \\
    & &{} \times e^{-2\pi i \sbvec{u}_k\cdot(\sbvec{x}_c-\sbvec{x}_k)}\,
    e^{2\pi i \sbvec{u}_{k'}\cdot(\sbvec{x}_{c'}-\sbvec{x}_{k'})}
\end{eqnarray}
where $\langle S_c(\nu_k)\,S_{c'}(\nu_{k'}) \rangle$ is the flux density
covariance matrix between sources $c$ and $c'$ at frequencies $\nu_k$
and $\nu_{k'}$ respectively.  There is a similar covariance matrix
$\langle V^{\rm src}_{k}\,V^{\rm src}_{k'} \rangle$.  T
These can be passed through the gridding procedure using equation
(\ref{eq:linop}) to make
\begin{equation}
  \bvec{\Delta}^{\rm src} = {\bf Q}\,\bvec{V}^{\rm src} + 
    \overline{\bf Q}\,\bvec{V}^{\rm src \ast}
\end{equation}
and used to construct the covariance elements
\begin{equation}
  {\bf M}^{\rm src} 
    = \langle \bvec{\Delta}^{\rm src}\,\bvec{\Delta}^{\rm src\,\dag} \rangle 
  \hspace{2cm}
  \overline{\bf M}^{\rm src} 
    = \langle \bvec{\Delta}^{\rm src}\,\bvec{\Delta}^{\rm src\,\rm t} \rangle
\end{equation}
using equation (\ref{eq:mmbar}).

This covariance matrix can be greatly simplified if we can subtract off the
mean source flux densities, leaving a zero-mean residual error.  
Let the true source flux density $S_c(\nu)$ be the sum of the measured flux
density $S^{\rm obs}_c(\nu)$ and an error $\delta S_c(\nu)$.  If our
measurements of these foreground sources are accurate, then the residuals
$\delta S_c(\nu)$ should be uncorrelated between sources (they are due to
measurement errors) and have zero mean.  In this case,
we can make corrected visibilities $V^{cor}_k$
\begin{equation}
  V^{cor}_k = V_k - \sum_c V^{\rm obs}_{ck}
  = \sum_c S^{\rm obs}_c(\nu_k)\,{A}_k(\bvec{x}_c-\bvec{x}_k)\,
  e^{-2\pi i \sbvec{u}_k\cdot(\sbvec{x}_c-\sbvec{x}_k)}
\end{equation}
to be used in place of $\bvec{V}$ in subsequent analysis.  Then, we are
left with the fluctuating component 
\begin{eqnarray}
  \delta V^{\rm src}_k & = & V^{\rm src}_k - \sum_c V^{\rm obs}_{ck}
  \nonumber \\
  \delta V_{ck} & = & \delta S_c(\nu_k)\,{A}_k(\bvec{x}_c-\bvec{x}_k)\,
  e^{-2\pi i \sbvec{u}_k\cdot(\sbvec{x}_c-\sbvec{x}_k)}
\end{eqnarray}
which we must deal with statistically.
The covariance between the source error contributions to the visibilities,
assuming the flux density errors are independent between sources (but not 
between frequency channels for the same source), is given by
\begin{eqnarray}
  \langle \delta V^{\rm src}_k \delta V^{\rm src \ast}_{k'} \rangle
    & = & \sum_c \langle \delta S_c(\nu_k) \delta S_{c}(\nu_{k'}) \rangle\,
    A_{k}(\bvec{x}_c-\bvec{x}_k)\,A^\ast_{k'}(\bvec{x}_c-\bvec{x}_{k'})\,
  \nonumber \\
    & &{} \times e^{-2\pi i \sbvec{u}_k\cdot(\sbvec{x}_c-\sbvec{x}_k)}\,
    e^{2\pi i \sbvec{u}_{k'}\cdot(\sbvec{x}_c-\sbvec{x}_{k'})}
\end{eqnarray}
and similarly for 
$\langle \delta V^{\rm src}_{k}\,\delta V^{\rm src}_{k'} \rangle$.
Finally, if the covariance is separable, e.g.
\begin{equation}
  \langle \delta S_c(\nu) \delta S_{c}(\nu') \rangle =
  \sigma_{Sc}(\nu)\,\sigma_{Sc}(\nu')
\end{equation}
then we can write
\begin{eqnarray}
  \langle \delta V^{\rm src}_k \delta V^{\rm src \ast}_{k'} \rangle & = &
    \sum_c \sigma^{\rm src}_{ck}\,\sigma^{\rm src \ast}_{ck'}
  \nonumber \\
  \sigma^{\rm src}_{ck} & = & \sigma_{Sc}(\nu_k)\,
    A_{k}(\bvec{x}_c-\bvec{x}_k)\,
    e^{-2\pi i \sbvec{u}_k\cdot(\sbvec{x}_c-\sbvec{x}_k)}.
  \label{eq:vsrcfluc}
\end{eqnarray}
The other covariance 
$\langle \delta V^{\rm src}_{k} \delta V^{\rm src}_{k'} \rangle$
can be computed in the same way. Because we have assumed that the
covariance is separable, we can speed up the covariance calculation as 
only the vector $\bvec{\sigma}^{\rm src}_{c}$ for each source is needed.
We can grid this onto the estimators
\begin{equation} 
  \bvec{\Delta}^{\rm src}_{c} = {\bf Q}\,\bvec{\sigma}^{\rm src}_c + 
    \overline{\bf Q}\,\bvec{\sigma}^{\rm src \ast}_c
  \label{eq:dsrcrms}
\end{equation}
and then
\begin{equation}
  {\bf M}^{\rm src} =  
    \sum_c \bvec{\Delta}^{\rm src}_{c}\,\bvec{\Delta}^{\rm src\,\dag}_{c} 
  \hspace{2cm}
  \overline{\bf M}^{\rm src} = 
    \sum_c \bvec{\Delta}^{\rm src}_{c}\,\bvec{\Delta}^{\rm src\,\rm t}_{c}
  \label{eq:mkksrc}
\end{equation}
which are used to build ${\bf C}^{\rm src}$.

There are two components to the source flux density uncertainties 
$\sigma_{Sc}(\nu_k)$, one from the uncertainties on the source
frequency spectrum, and the other from the uncertainties on the flux density
measurements and any extrapolation of the measured flux densities to the
observing frequencies $\nu_k$ (using the estimated source spectrum).
As an example, consider a source with a flux density $S_c(\nu_0)$ measured
with standard deviation $\sigma_{S0}$ at frequency $\nu_0$, and a known
power-law frequency spectrum with spectral index $\alpha$, 
\begin{equation}
  S_c(\nu_k) = S_c(\nu_0)\,f(\nu_k/\nu_0,\alpha) \hspace{2cm}
  f(\nu_k/\nu_0,\alpha) = \left( {\nu_k \over \nu_0} \right)^\alpha.
  \label{eq:fnunu0}
\end{equation}
Then, it is easy to show that
\begin{equation} \label{eq:sigmas}
  \sigma_{Sc}(\nu_k)/S_c(\nu_k) = \sigma_{S0}/S_c(\nu_0)
\end{equation}
with the fractional uncertainty in the flux density $\sigma_{Sc}/S_c$
remaining independent of the frequency.

On the other hand, consider the case where there is now an uncertainty
$\sigma_\alpha$ in the spectral index between $\nu_k$ and $\nu_0$.
Then, our extrapolation factor $f(\nu/\nu_0,\alpha)$, which we write as 
\begin{equation}
  f(\nu/\nu_0,\alpha) = e^{\alpha\,\ln(\nu/\nu_0)},
\end{equation}
propagates to the extrapolated flux density as
\begin{equation} \label{eq:sigmaa}
  \sigma_{Sc}(\nu_k)/S_c(\nu_k) = \ln( \nu/\nu_0 )\,\sigma_\alpha
\end{equation}
which can be negative --- for two channels flanking the fiducial 
frequency (e.g.\ $\nu < \nu_0 < \nu'$) the errors will be anti-correlated.
Note that we have approximated the resulting distribution as Gaussian.
In general it is not, e.g.\ for a Gaussian distribution in $\alpha$ we
find a log-normal distribution in $S(\nu)$.

Although the dominant spectral error is due to the extrapolation from
a frequency $\nu_0$ outside the range of the CMB instrument, there is an
additional 
error due to an error in the spectral index over the frequency channels
$\nu_k$ of the visibilities.  This is as if you extrapolated using one
spectrum appropriate for the band center $\bar{\nu}$ of the instrument,
but when the flux densities $S^{\rm obs}_c(\nu_k)$ are extrapolated from band
center $S^{\rm obs}_c(\bar{\nu})$ there is an error from using the wrong
$\alpha$ over the band.  This is handled using equation (\ref{eq:sigmaa}) 
with another $\sigma_\alpha$ appropriate to the uncertainty in the spectral
index over the $\nu_k$.

For the CBI analysis, we have approximated both the flux density error
and the spectral extrapolation error as a single equivalent flux density
error.  For the CBI, the frequency span (26--36~GHz, or $d\nu/\nu=\pm 16\%$)
is small enough that we can approximate the spectral index uncertainty as
an effective flux density uncertainty $\sigma_c$ extrapolated to band center
$\bar{\nu}$ from $\nu_0$ using $\alpha_0$
\begin{equation} \label{eq:sigmeff}
  \sigma_{Sc}(\nu_k) \approx 
  f(\nu/\bar{\nu},\alpha)\,\sigma_{c} \hspace{2cm}
  \sigma^2_{c} = f^2(\bar{\nu}/\nu_0,\alpha_0)\,\Big( \sigma^2_{S0} + 
     S^2_0\,\big[ \ln( \bar{\nu}/\nu_0 ) \big]^2\,\sigma^2_\alpha \Big)
\end{equation}
where $\alpha$ need not equal $\alpha_0$ and should reflect the spectral
index over the observing band, not the one used for extrapolation from 
$\nu_0$. 

In principle, if the true mean flux densities for the sources are correctly
subtracted from the visibilities, and the covariance matrix ${\bf C}^{\rm src}$
is built using the correct elements
$\langle \delta S_c(\nu_k) \delta S_{c'}(\nu_{k'}) \rangle$, then
inclusion of this as a noise term in ${\bf C}$ using $q_{\rm src}=1$ would
remove the effects of these sources from our power spectrum estimation in a 
statistical sense.  However, there are a number of factors that make this
difficult.  If the source flux density measurements have a calibration error,
then the errors will not be independent between sources.  Also, the fainter
sources (which are still significant contributors to the signal) have
flux densities that are often extrapolated from much lower frequencies 
(e.g.\ the ``NVSS'' sources in Paper~II and Paper~III).  Furthermore, since
there are a relatively small number of discrete sources contributing to a
given field, it is not clear we are in the statistical limit where the
flux density covariance is an accurate description of what is happening to
the data.  For these reasons, for the CBI analysis we treat the covariance
matrix ${\bf C}^{\rm src}$ constructed using the approximation in equation 
(\ref{eq:sigmeff}) as a constraint matrix for the nuisance parameters
due to the sources, and set $q_{\rm src}$ to a high enough amplitude to 
{\it project out} the contaminated modes in the data.  Because the source
modes are spread out by the effect of the synthesized beam (the ``point-spread
function'' in imaging terms), setting $q_{\rm src}$ to too high a value will
start to down-weight modes that are not significantly contaminated, while too
low a value will eat into the noise and CMB signal power in those modes
without down-weighting them sufficiently thus biasing the affected bandpowers
low.  The exact values to be used thus depend upon the signal and noise
levels in the data; we refer the reader to Papers II and III for descriptions
of what was chosen for the CBI analysis.  See \citet{bjk98} for a description
of the constraint matrix formalism and the technique of projection.

\subsection{Gaussian Foregrounds and Residual Point Sources}
\label{sec:foreground}

In \S\ref{sec:pscmb} it was mentioned that a single foreground
component could be modeled with a modified covariance matrix, power spectrum
shape and frequency dependence. As long as these foregrounds can be treated as
a Gaussian random field, they can be processed in the same manner as the CMB.
Therefore, once the amplitude and shape of the foreground fluctuation power
spectrum $C_{\rm res}(v)$ is input, we compute the foreground covariance matrix
elements 
\begin{equation}
   {\bf M}^{\rm res} = \sum_\ell { {\bf W}^{\rm res}_{\ell} \over \ell }\,
   {\cal C}^{\rm res}_\ell \,  
   \hspace{2cm}
   \overline{\bf M}^{\rm res} = \sum_\ell 
   { \overline{\bf W}^{\rm res}_{\ell} \over \ell }\,{\cal C}^{\rm res}_\ell
   \label{eq:mkkfg}
\end{equation}
where the variance window functions ${\bf W}^{\rm res}$ and 
$\overline{\bf W}^{\rm res}$ are given by substituting for 
${\bf R}$ in equation (\ref{eq:wijvarpi}) a new ${\bf R}^{\rm res}$ built
from a kernel
\begin{equation} 
  P^{\rm res}_k(\bvec{v}) = f^{\rm res}_k\,\tilde{A}_k(\bvec{u}_k-\bvec{v})\,
  e^{2\pi i \sbvec{v}\cdot\sbvec{x}_k}
  \label{eq:pkres}
\end{equation}
using a frequency factor $f^{\rm res}_k=f^{\rm res}(\nu_k)$ appropriate to the
foreground in question.  The matrix ${\bf C}^{\rm res}$ is
then obtained by substitution of 
${\bf M}^{\rm res}$ and $\overline{\bf M}^{\rm res}$ 
as usual using equation (\ref{eq:csbreimest}).
Although it is possible to break up the Gaussian foreground component into
bands as we did the CMB, it is preferable to compute the foreground
covariance matrix in a single band using its shape ${\cal C}^{\rm res}_\ell$, 
to reduce the degeneracy with the CMB --- you cannot distinguish between the
two in narrow $\ell$ bands where the shapes are unimportant.

An example of a foreground that strongly affects the CBI data is that from
point sources below the limit for subtraction contaminating the CBI fields.
This {\it residual} statistical background, in the limit where there are many
sources per field, can be modeled as a white noise Gaussian field with
constant angular power spectrum and power-law frequency spectrum.  Each
individual source has a flux density drawn from a differential number count
distribution $dN(S_\nu)/dS$ which represents the number of sources per
steradian with flux densities between $S_\nu$ and $S_\nu+dS$ at observing
frequency $\nu$. The angular clustering in these sources is very small, and
can be neglected. 

The contribution of a source $c$ to visibility $V_k$ was given by 
$V_{ck}$ in equation (\ref{eq:vsrc}).  The sources are independently
distributed in flux density and on the sky, so
\begin{eqnarray}
   \langle V_{k}\,V^\ast_{k'} \rangle & = &
   \langle \sum_c S_c(\nu_k)\,S_{c}(\nu_{k'})\,
   A_{k}(\bvec{x}_c-\bvec{x}_k)\,A^\ast_{k'}(\bvec{x}_c-\bvec{x}_{k'})\,
   e^{2\pi i \sbvec{u}_k\cdot(\sbvec{x}_c-\sbvec{x}_k)}
   e^{-2\pi i \sbvec{u}_{k'}\cdot(\sbvec{x}_c-\sbvec{x}_{k'})} \rangle 
   \nonumber \\
   & = & {1 \over \Omega}\,
   \langle \sum_c S_c(\nu_k)\,S_{c}(\nu_{k'}) \rangle \,B_{kk'} 
   \nonumber \\
   & = & {C}^{\rm res}(\nu_k,\nu_{k'})\,B_{kk'}
\end{eqnarray}
where the angular average can be written as an integral over 
\begin{equation}
   B_{kk'} = \int d^2\bvec{x}\,
   A_{k}(\bvec{x}-\bvec{x}_k)\,A^\ast_{k'}(\bvec{x}-\bvec{x}_{k'})\,
   e^{2\pi i \sbvec{u}_k\cdot(\sbvec{x}-\sbvec{x}_k)}\,
   e^{-2\pi i \sbvec{u}_{k'}\cdot(\sbvec{x}-\sbvec{x}_{k'})} 
\end{equation}
with $\Omega$ as the normalizing solid angle.
This integral is just a Fourier transform, and so
\begin{equation}
   B_{kk'} =  \int d^2\bvec{v}\,
   \tilde{A}_{k}(\bvec{u}_k-\bvec{v})\,
   \tilde{A}^\ast_{k'}(\bvec{u}_{k'}-\bvec{v})\,
   e^{2\pi i \sbvec{v}\cdot(\sbvec{x}_k-\sbvec{x}_{k'})}
\end{equation}
which is the same as the CMB visibility covariance matrix $M_{kk'}$ in
equation (\ref{eq:mkkintvis}) with $f_k=f_{k'}=1$ and $C(v)=1$.  Similarly,
the other 
covariance $\langle V_{k}\,V_{k'} \rangle$ reduces to $\overline{M}_{kk'}$.
Thus, in the stochastic limit the residual source background behaves as a
Gaussian random field and can thus be treated as we do the CMB signal in 
\S\ref{sec:viscor} but with a power spectrum shape 
$C_\ell={C}^{\rm res}(\nu_k,\nu_{k'})$ which is constant over $\ell$ for a
given pair of frequency channels.

The amplitude of the covariance matrix is the ensemble average of the 
source power per solid angle, which is obtained by integration over the flux
density and spectral index distributions
\begin{equation} \label{eq:resampint}
  {C}^{\rm res}(\nu_k,\nu_{k'}) = 
    \int_{S_{\rm min}}^{S_{\rm max}} dS\,S^2\,{ dN(S) \over dS }\,
    \int_{-\infty}^{\infty} d\alpha\,p(\alpha|S,\nu_0)\,
    \left( {\nu_k\,\nu_{k'} \over \nu^2_0} \right)^\alpha
\end{equation}
where we have again assumed that the spectrum is a power-law with spectral
index $\alpha$ over the range of interest for the $\nu_k$, and integrate
over the number counts over the flux density range from 
$S_{\rm min}$ to $S_{\rm max}$.  We also assume that there is a large number
of these faint sources over the solid angles of interest (e.g.\ the 
CBI primary beam) and thus the Poisson contribution to the 
probability can be ignored and we can use the mean source density given
by the number counts $dN/dS$ at the fiducial frequency $\nu_0$ for which
$S$ is given.  The spectral index distribution as a function of flux density
$p(\alpha|S,\nu_0)$ must be estimated from radio surveys, though it can
be uncertain at the high frequencies and faint levels at which the CMB
experiments are carried out.  If $p(\alpha|S,\nu_0)=p(\alpha|\nu_0)$ and
thus is independent of flux density, then it can be shown 
(e.g.\ Appendix~\ref{app:counts}) that the
integrals in equation (\ref{eq:resampint}) can be evaluated at a single
frequency $\nu$ in the band and scaled using an effective spectral index
$\alpha_{\rm eff}$
\begin{equation} \label{eq:creskk}
  {C}^{\rm res}(\nu_k,\nu_{k'}) = {C}^{\rm res}_\nu\,
  f^{\rm eff}_k\,f^{\rm eff}_{k'}
  \hspace{2cm}
  f^{\rm eff}_k = \left( {\nu_k \over \nu} \right)^{\alpha_{\rm eff}}
\end{equation}
where ${C}^{\rm res}_\nu$ is the amplitude of the fluctuation power per solid
angle (in units of Jy$^2$/sr) at $\nu$.  In terms of the logarithmic
average ${\cal C}$ for the CMB, 
${\cal C}^{\rm res}_\nu=\ell^2\,{C}^{\rm res}_\nu/2\pi$
which rises at high $\ell$ with respect to the CMB.
See Appendix~\ref{app:counts} for an example analytic
calculation using power-law source counts and a Gaussian spectral index
distribution.

The frequency range of the CBI is insufficient to distinguish nonthermal
foreground emission from the thermal CMB, and thus this is treated as a
constraint matrix (i.e.\ $q_{\rm res}$ is not solved for as a parameter).
Therefore, in the CBI analysis we construct the covariance matrix ${\bf
C}^{\rm res}$ using the matrix elements in equation (\ref{eq:mkkfg}) built
assuming unit 
power (1~Jy$^2$/sr) and the frequency dependence $f_k=f^{\rm eff}_k$ from
equation (\ref{eq:creskk}).   The value used for $q_{\rm res}$ is equal to the
source fluctuation power ${\cal C}^{\rm res}_\nu$  calculated as an
{\it a priori} estimate based on knowledge of the residual foreground source
populations (see Appendix~\ref{app:counts}).  

\subsection{Other Signal Components}\label{sec:othersrc}

We are not restricted to CMB, Gaussian foreground, and discrete point
sources as the components of our signal or noise in the covariance matrix
${\bf C}$ in equation (\ref{eq:csnbandv}).  This approach can be generalized
to deal with other signals of interest.  For example, extended sources with
a known profile, such as the Sunyaev-Zeldovich Effect from clusters
of galaxies, could be modeled either analytically or numerically giving
a power spectrum shape (e.g.\ \citealt{bm96}).  In the case of a signal
with a known distribution on the sky, a template can be used.  Examples
of this include dust emission in the millimeter-wave bands, or the anomalous
centimeter-wave emission observed at OVRO \citep{ovro5m} and by COBE
\citep{kogut}.  In particular, the latter foreground, which is posited as due
to spinning dust grains by \citet{dl99}, correlates with the 100$\mu$m dust
emission as measured by IRAS and DIRBE, and thus a template of emission can be
constructed.

\subsection{Differencing}\label{sec:difference}

Unfortunately, with its low intrinsic fringe rates and extremely short
($<90\lambda$) spacings, the CBI is susceptible to ground pickup.  
To remove this, we observe for each field a trailing 
field displaced $8^{\rm m}$ in right ascension $8^{\rm m}$ later, and
difference the corresponding visibilities.  Therefore, we must take this
differencing into account in our correlation analysis.  This effectively
modifies the window function, quenching low spatial frequencies further.

Let us write
\begin{equation} \label{eq:vxsw}
  V^{\rm sw}_k = V^{\rm main}_k - V^{\rm trail}_k \hspace{2cm}
  \bvec{x}^{\rm trail}_k = \bvec{x}_k + \Delta\bvec{x}_k
\end{equation}
for switching offset $\Delta\bvec{x}_k$ (e.g.\ $8^{\rm m}$ in RA $\approx
2^\circ$ for the CBI fields near the celestial equator).
Then, from equation (\ref{eq:visi}) we find
\begin{equation} \label{eq:vksw}
  V^{\rm sw}_{k} = f_k\,\int d^2\bvec{v}\, \tilde{A}_k(\bvec{u}_k-\bvec{v})\,
  \tilde{T}(\bvec{v}) \, e^{2\pi i \sbvec{v}\cdot\sbvec{x}_k} \,
  \left[ 1 - e^{2\pi i \sbvec{v}\cdot\Delta\sbvec{x}_k} \right]
   + e^{\rm sw}_k
\end{equation}
where the switched noise $e^{\rm sw}_k=e^{\rm main}_k - e^{\rm trail}_k$.
In terms of the kernel of equation (\ref{eq:pk}),
\begin{equation}
  V^{\rm sw}_{k} = \int d^2\bvec{v}\, P^{\rm sw}_k(\bvec{v})\,
  \tilde{T}(\bvec{v}) + e^{\rm sw}_k
  \hspace{2cm}
  P^{\rm sw}_k(\bvec{v}) = P_k(\bvec{v})\,
  \left[ 1 - e^{2\pi i \sbvec{v}\cdot\Delta\sbvec{x}_k} \right]
  \label{eq:pksw}
\end{equation}
and thus for our switched visibilities we compute everything as before, but
substituting $P^{\rm sw}_k$ for $P_k$.

Note that if the trail field offsets $\Delta\bvec{x}$ were constant in
arc length rather than in Right Ascension (this is approximately true since
the declination range of the mosaic is limited) we could write the convolution
kernel as
\begin{equation} \label{eq:ppsw}
  P^{\rm sw}_k(\bvec{v})\,P^{\rm sw\ast}_{k'}(\bvec{v}) = 
  P_k(\bvec{v})\,P^\ast_{k'}(\bvec{v}) \,
  \left[ 2 - 2\,\cos( 2\pi \bvec{v}\cdot\Delta\bvec{x}) \right]
\end{equation}
where the leading factor of 2 dominates (you essentially get twice the
CMB power).  Note that a noticeable effect of the differencing is that
the window function will have a ripple of ``wavelength'' $\Delta x^{-1}$
superimposed on its envelope.  For example, the $8^{\rm m}$ switching
in RA that the CBI uses corresponds to $\Delta x=2^\circ$ at the celestial
equator, and thus the ripple has $\Delta x^{-1}=28.6$ in $u$.  This
corresponds to 180 in $\ell$, but is azimuthally averaged in the 
{\it uv}-plane and thus the peak-to-peak amplitude is reduced.

\section{Solving the Likelihood Equation} \label{sec:relax}

We have expressed the estimators as a real vector
$\bvec{d}$ and obtained expressions for the components of its covariance
matrix, and now turn to the problem of solving
for the maximum likelihood estimators for the bandpowers using equation 
(\ref{eq:likx}).  As shown below, we will be carrying out a large number
of matrix operations using ${\bf C}$ and its component matrices 
(e.g.\ ${\bf C}^{\rm N}$, ${\bf C}^{\rm S}_B$, etc.) and thus these will need
factorization.   Because ${\bf C}$ is positive definite
we use optimized Cholesky decomposition routines (DCHDC from LINPACK, or
DPOTRF from LAPACK)\footnote{available from \tt http://www.netlib.org/}
to carry out the required factorizations.

The large number of visibilities {\it times} the number of
mosaic  pointings makes this computation extremely costly (the 
matrix inversions and/or solution of systems of equations are order $N^3$
processes!), especially for a large number of 
bands $N_B$.  Clever perturbative or gradient search
methods can help to reduce the overhead in finding the maximum in
parameter space.
One such method is the quadratic relaxation technique of \citet[BJK]{bjk98}.
To summarize here, if one Taylor expands 
the log-likelihood around the maximum likelihood bandpowers 
$\hat{\bvec{q}}=\{\hat{q}_B, B=1,\ldots,N_B\}$ to second order
\begin{equation}
  \ln L(\hat{\bvec{q}} + \delta\bvec{q}) = \ln L(\hat{\bvec{q}}) + 
  \sum_B {\partial \,\ln L(\hat{\bvec{q}}) \over \partial q_B }
  \,\delta q_B + 
  {1\over2}\,\sum_{B\phantom{'}} \sum_{B'\phantom{'}} 
  {\partial^2 \,\ln L(\hat{\bvec{q}}) \over \partial q_B\,\partial q_{B'} }
  \,\delta q_B\,\delta q_{B'}
\end{equation}
then we can move toward the maximum using the quadratic approximation
\begin{equation}
  \delta q_B = - \sum_{B'}
  \left[ {\partial^2 \,\ln L({\bvec{q}}) \over \partial q_B\,\partial q_{B'} }
  \right]^{-1} \,
  {\partial \,\ln L({\bf q}) \over \partial q_{B'} }.
\end{equation}

The first derivative (gradient) is given by
\begin{equation}
  {\partial \,\ln L(\bvec{q}) \over \partial q_{B'}} = 
  {1\over2}\,{\rm Tr}\left[ (\bvec{d}\,\bvec{d}^{\rm t} - {\bf C})
  ( {\bf C}^{-1}\,{\partial C \over \partial q_{B'}}\,
    {\bf C}^{-1} ) \right]
\end{equation}
and the second derivative (curvature matrix) by
\begin{eqnarray}
  {\cal F}_{B B'} & = & 
  - {\partial^2 \,\ln L(\bvec{q}) \over \partial q_B\,\partial q_{B'} } 
  \nonumber \\
  & = &
  {\rm Tr}\bigg[ (\bvec{d}\bvec{d}^{\rm t} - {\bf C})
  ( {\bf C}^{-1}\,{\partial C \over \partial q_{B}}
    {\bf C}^{-1}\,{\partial C \over \partial q_{B'}}\,{\bf C}^{-1}
  - {1\over2}{\bf C}^{-1} 
    {\partial^2 C \over \partial q_{B}\,\partial q_{B'}}
    {\bf C}^{-1} )
   \bigg] \nonumber \\ 
  & &{} + {1\over2}\,{\rm Tr}\bigg[ 
    {\bf C}^{-1} {\partial C \over \partial q_{B}}
    {\bf C}^{-1} {\partial C \over \partial q_{B'}}
    \bigg].
  \label{eq:curvf}
\end{eqnarray}
Note that the partial derivatives of the covariance matrix are just
the band signal covariance matrices 
$\partial {\bf C}/\partial q_{B} = {\bf C}^{\rm S}_B$ defined above.

The final approximation is to replace the curvature matrix with its
expectation value, which is the Fisher information matrix
\begin{equation}
   F_{B B'} = \left\langle {\cal F}_{B B'} 
   \right\rangle =
   {1\over2}\,{\rm Tr}\left[ 
   {\bf C}^{-1}\,{\bf C}^{\rm S}_B\,{\bf C}^{-1}\,{\bf C}^{\rm S}_{B'} 
   \right]. \label{eq:fisher}
\end{equation}
This yields
\begin{equation} 
  \delta q_B = {1\over2}\,\sum_{B'}
  \left[ F^{-1} \right]_{B B'} \, 
  {\rm Tr}\left[ (\bvec{d}\,\bvec{d}^{\rm t} - {\bf C})
  ( {\bf C}^{-1}\,{\bf C}^{\rm S}_{B'}\,{\bf C}^{-1} ) \right] \label{eq:delqb}
\end{equation}
for the iterative correction to the bandpowers.  This amounts to
making a quadratic approximation to the shape of the likelihood function
around the maximum, and iteratively approaching it.
At each step, the total covariance matrix ${\bf C}$ must be updated using
the new bandpowers $\bvec{q}+\delta\bvec{q}$.  A convergence criterion
based on the magnitude of the corrections $\delta q_B$ will allow approach to
the true $\{\hat{q}_B\}$ to be controlled.

The inverse of the Fisher matrix $[ F^{-1} ]_{B B'}$ evaluated at
maximum likelihood is the covariance matrix of the parameters \citep{bjk98}.
The diagonals $[ F^{-1} ]_{BB}$ give an estimated Gaussian error bar
for the derived bandpowers $\{\hat{q}_B\}$, though the full Fisher matrix
must be used to take the (usually significant) band-band correlations into
account.  As the width of the $\ell$ bins for the bands $B$ is reduced, 
anti-correlation between adjacent bands increases due to the intrinsic
$\ell$-space resolution of the data.

The presence of known or residual point source foregrounds in equation
(\ref{eq:csnbandv}) is dealt with by either fixing the amplitudes
$q_{\rm src}$ or $q_{\rm res}$ and treating $q_{\rm src}\,{\bf C}^{\rm src}$
or $q_{\rm res}\,{\bf C}^{\rm res}$ as additions to the noise matrix 
${\bf C}^{\rm N}$, or by solving for the $q_{\rm src}$ or $q_{\rm res}$ and
treating them as extra bandpowers $q_B$ with associated entries in the Fisher
matrix. In practice, for the CBI, it is necessary to hold fixed the 
$q_{\rm res}$ because the contribution from the source foreground with a white
noise power spectrum and appropriate frequency spectrum is largely
indistinguishable from an overall offset of the CMB power spectrum.
In addition, the uncertainties on the individual
known source contributions to an aggregate ${\bf C}^{\rm src}$ will be
substantial, and thus solving for a single amplitude $q_{\rm src}$ will not
be as useful as it might appear.  In this case, the ${\bf C}^{\rm src}$
acts as a constraint matrix and the $q_{\rm src}$ can be
set to an arbitrarily high value, which will effectively {\it project}
out the modes corresponding to the known sources by down-weighting the
relevant combinations of the estimators in the likelihood \citep{bjk98,bc01}.

\subsection{Combination of Independent Datasets}\label{sec:comblike}

Consider observations taken of separate sets of single fields or 
mosaics $f$ where there
is effectively no correlation between fields from separate $f$ and
the fields within a given set $f$ are related by the mosaic covariance
given in the previous sections.  In this case, we can assemble a giant
data vector
\begin{equation} \label{eq:xfx}
  \bvec{D} =
  \left( \begin{array}{ccc} \bvec{d}_1 & \ldots & 
  \bvec{d}_M \end{array} \right)^{\rm t}
\end{equation}
from the $M$ individual field vectors $\bvec{d}_f$ (e.g.\ eq.[\ref{eq:xvec}]),
with the block diagonal covariance matrix
\begin{equation} \label{eq:cfx}
  {\bf C} =
  \left( \begin{array}{ccc} 
         {\bf C}_1 & & \\
         & \ldots & \\
         & & {\bf C}_M \end{array} \right)
\end{equation}
which in turn can be written as sums of block-diagonal noise and signal
covariance matrices ${\bf C}^{\rm N}_f$ and ${\bf C}^{\rm S}_{Bf}$ etc., with
blocks given by ${\bf C}^{\rm N}_f$ and ${\bf C}^{\rm S}_{Bf}$, etc. Because
they are block diagonal, we can write the log-likelihood in equation
(\ref{eq:likx}) as the sum over datasets
\begin{eqnarray}
  \ln L & = & -\ln(2\pi)\,\sum_f  N_f - {1\over2}\,\ln ( \det {\bf C} ) - 
  {1\over2}\,\bvec{D}^{\rm t}\,{\bf C}^{-1}\,\bvec{D} \nonumber \\
  & = & -\ln(2\pi)\,\sum_f  N_f - {1\over2}\,\sum_f \ln ( \det {\bf C}_f ) - 
  {1\over2}\,\sum_f \bvec{d}_f^{\rm t}\,{\bf C}_f^{-1}\,\bvec{d}_f.
\end{eqnarray}

We proceed as before, with the same bandpowers $\{q_B\}$ and with
the block-diagonal band covariance matrix
\begin{equation} \label{eq:csfx}
  {\bf C}^{\rm S}_B = {\partial {\bf C} \over \partial q_{B}} = 
  \left( \begin{array}{ccc} 
         {\bf C}^{\rm S}_{B1} & & \\
         & \ldots & \\
         & & {\bf C}^{\rm S}_{BM} \end{array} \right)
\end{equation}
and thus all matrices are block-diagonal and composed of the individual
single field or mosaic matrices.  Therefore,
\begin{eqnarray}
  F_{B B'} & = & {1\over2}\,\sum_f {\rm Tr}\left[ 
    {\bf C}_f^{-1}\,{\bf C}^{\rm S}_{Bf}\,{\bf C}_f^{-1}
    \,{\bf C}^{\rm S}_{B'f} \right] \nonumber \\
  \delta q_B & = & {1\over2}\,\sum_{B'}\left[ F^{-1} \right]_{B B'} \, 
    \sum_f {\rm Tr}\left[ (\bvec{d}_f\,\bvec{d}_f^{\rm t} - {\bf C}_f)
    ( {\bf C}_f^{-1}\,{\bf C}^{\rm S}_{B'f}\,{\bf C}_f^{-1} ) \right]
\end{eqnarray}
which is used to iteratively approach the $\{\hat{q}_B\}$ using the BJK scheme
as in the single dataset case.

\subsection{The Bandpower Window Function} \label{sec:bwin}

To compare the bandpowers obtained from the data to model
power spectra we need to define a set of filter functions which
project models ${\cal C}_{\ell}$ into a set of bandpowers $C_B$
\begin{equation}\label{eq:bpaverage}
   C_B  = {\sum_{\ell} { W^B_{\ell} \over \ell}\,\cal{C}_{\ell}}
\end{equation}
as in \citet{bjk98}.
In the ensemble limit, the expectation value 
$\langle (\bvec{x}\,\bvec{x}^{\rm t} - {\bf C}^{\rm N}) \rangle$ 
will approach the underlying signal covariance matrix ${\bf C}^{\rm S}$. We
can then use the expression for the minimum variance estimate of the bandpower
to derive the filter functions $W^B_{\ell}$ \cite{knox}. Since
\begin{equation}
  \langle q_B \rangle  = {1\over2}\,\sum_{B'}
  \left[ F^{-1} \right]_{B B'} \,
  {\rm Tr}\left[ ( {\bf C}^{-1}\,{\bf C}^{\rm S}_{B'}\,{\bf C}^{-1}) \,
  {\bf  C}^{\rm S} \right]
\end{equation}
and
\begin{equation}
  {\bf C}^{\rm S} \equiv \sum_{B} {\bf C}^{\rm S}_B =
  \sum_{\ell} \frac{\partial {\bf C}^{\rm S}}{\partial {\cal C}_{\ell}}
  \,{\cal C}_{\ell}
\end{equation}
the normalized filter functions can be computed using the band
averaged Fisher matrix (e.g.\ eq.[\ref{eq:fisher}]) 
\begin{equation}\label{eq:windows}
  { W^B_{\ell} / \ell } = {1\over2}\,\sum_{B'}
  \left[ F^{-1} \right]_{B B'} \,
  {\rm Tr}\left[ ( {\bf C}^{-1}\,{\bf C}^{\rm S}_{B'}\,{\bf C}^{-1})\,
  \frac{\partial {\bf C}^{\rm S}}{\partial \cal{C}_{\ell}}\right]
\end{equation}
with respect to the ${\cal C}^{\rm shape}_{\ell} = 1$ that is built into
the ${\bf C}^{\rm S}$.  Because of the $\chi_{B\ell}$
used in the construction of the ${\bf C}^{\rm S}_B$ in equation
(\ref{eq:mkkbin}), 
\begin{equation} \label{eq:winortho}
   \sum_{\ell} \chi_{B'\ell}\,W^B_{\ell}/\ell = \delta_{BB'}
\end{equation}
and thus $W^B_{\ell}/\ell$ is orthonormal with respect to the bands
defined by $\chi_{B\ell}$.

Calculating the filter functions at each $\ell$ becomes somewhat
prohibitive in both
processor time (the problem scales as an extra $N^3 + 2\ell_{\rm max}N^2$
operations) and storage since the calculation of equation (\ref{eq:windows})
can only proceed once we have relaxed to the maximum likelihood solution.
For this reason, in practice we sample the full filter functions in bands
at intervals $B_f$ where the $B_f$ are narrower than the bands $B$, with
\begin{equation} \label{eq:wbbf}
  { W^B_{B_f} / \ell_{B_f} } =
  {1\over2}\,\sum_{B'}\left[ F^{-1} \right]_{B B'} \,
  {\rm Tr}\left[ ( {\bf C}^{-1}\,{\bf C}^{\rm S}_{B'}\,{\bf C}^{-1}) \,
  {\bf C}^{\rm S}_{B_f}\right].
\end{equation}
In principle, this is equivalent to 
assuming a flat window over the `fine' band $B_f$ and as long as the
curvature of the exact window function is small over the intervals $B_f$
this should provide an adequate sampling of the continuous limit.

To obtain model bandpowers we can then either interpolate the samples
$W^B_{B_f}$ to obtain an approximate form for $W^B_{\ell}$ 
for use in equation (\ref{eq:bpaverage}) or {\it pre-average} the model
spectrum over the fine bands $B_f$ as
\begin{equation} \label{eq:cbpred}
   C_B = 
   \sum_{B_f} (W^B_{B_f}/\ell_{B_f})\,{\cal C}^{({\rm flat})}_{B_f}
\end{equation}
where ${\cal C}^{({\rm flat})}_{B_f}$ are bandpowers calculated using flat
filters ($W^{B_f}_{\ell} = 1$). We find that a fine band width
$\Delta\ell_{B_f} \sim 20$ is sufficient to adequately sample the window
functions and ensure normality and orthogonality to within $0.5\%$ with
respect to integration over the bands (e.g.\ eq.[\ref{eq:winortho}]).
Example window functions calculated in this manner for mock deep fields and
mosaics are shown in the lower panels of Figure~\ref{fig:mockdeep} and
Figure~\ref{fig:mockmos} respectively.

\subsection{Component Bandpowers} \label{sec:comp}

A further complication at the parameter end of the process is that the
likelihood of the bandpowers cannot be assumed to be a Gaussian. This
is especially so in cases where the error in the bandpowers is sample
or cosmic variance limited. Assuming the bandpowers to be Gaussian
distributed can lead to the well known problem of {\it cosmic bias}
where the likelihood of low power models can be overestimated and
conversely that of high power models can be underestimated.  \citet{bjk00}
have shown how one can avoid this problem while still retaining
Gaussianity in the $\chi^2$ analysis by treating certain functions of
the bandpowers as Gaussian distributed. Very good fits to the
non-Gaussian distribution of the bandpowers can be obtained by use of
the {\it offset log-normal} and {\it equal variance} approximations to
the likelihood.

Both approximations use offsets $x_B$ in the bandpowers which describe
the contributions to the error in the bandpowers due to components
other than the CMB. For the range of scales probed by instruments such
as CBI these components will include the foregrounds such as point
sources in addition to the usual noise `on the sky' offset 
$x^{\rm N}_B\sim \sum_{\ell} \chi_{B\ell}\,x_{\ell}$,
where $x_\ell$ is the offset due to the noise contribution to the
error such that the quantity $Z_\ell=\ln({\cal C}_\ell + x_\ell)$ has a normal
distribution \citet{bjk00}.  For accurate parameter fits we therefore
require estimates of bandpowers for all the components making up the
total covariance ${\bf C}$. An approximation for these can be
obtained by modifying the minimum variance estimator for the
bandpowers $q_B$ at the maximum likelihood
\begin{equation}
  q_B^{\rm X}  = {1\over2}\,\sum_{B'}
  \left[ F_{B B'} \right]^{-1} \,
  {\rm Tr}\left[ ( {\bf C}^{-1}\,{\bf C}^{\rm S}_{B'}\,{\bf C}^{-1}) \,
  {\bf  C}^{\rm X} \right]
\end{equation}
where we have substituted ${\bf C}^{\rm X}$ in equation (\ref{eq:delqb}) for
the observed  measure for the signal
covariance $({\bf d}\,{\bf d}^{\rm t} - {\bf C}^{\rm N})$.  
We then set ${\bf C}^{\rm X}$
to the noise ${\bf C}^{\rm N}$, foreground source  ${q}_{\rm src}\,{\bf
C}^{\rm src}$ or Gaussian residual foreground ${q}_{\rm res}\,{\bf
C}^{\rm res}$ covariance components as desired (or use the maximum
likelihood values $\hat{q}_{\rm src}$ and $\hat{q}_{\rm res}$ if these
are included as parameters in the solution rather than being fixed).  
Examples of these are shown in Paper~II and Paper~III for the deep field data
and mosaic data respectively.  The offset to the log-normal is then obtained
by summing the $q_B^{\rm X}$ over the components, 
$x_B \approx q_B^{\rm N} + q_B^{\rm src} + q_B^{\rm res}$ 
(e.g.\ \citealt{bc01}).
This formalism is used in Paper~V to approximate the shapes of the likelihood
functions in order to derive limits on the cosmological parameters.

\section{Imaging from the Gridded Estimators} \label{sec:image}

Although not the primary goal of this method, an image can be constructed
by Fourier transforming back to the sky plane using equation 
(\ref{eq:fourier}).  If the estimators $\Delta_i$ are constructed on a regular
lattice in ${\bf u}_i$ with spacing $d_u$ and {\it uv} extent $L\,d_u$, 
then the resulting image
will have an extent on the sky given by the inverse of the spacing
$d_u^{-1}$ and a resolution given by $d_u^{-1}/L$.  In the
continuum limit (see Appendix~\ref{app:estim}), we can define an estimator
$\hat{T}(\bvec{x})$ for the temperature field ${T}(\bvec{x})$
\begin{equation} \label{eq:image}
   \hat{T}(\bvec{x}) = \int d^2\bvec{u}\, \Delta(\bvec{u})\,
   e^{2\pi i \sbvec{u}\cdot\sbvec{x}}
\end{equation}
where $\Delta(\bvec{u})$ is the continuous functional form 
(e.g.\ eq[\ref{eq:estcont}]) for the estimators, with
$\Delta_i=\Delta(\bvec{u}_i)$.
In practice the lattice of estimators $\Delta_i$ 
is embedded in a wider grid padded with zero in the unsampled cells,
and an FFT is carried out.

For our standard gridding normalization $z_i=z^{(1)}_i$
given in equation (\ref{eq:zk1}), the units of $\bvec{\Delta}$ will be 
flux density units (Jy) and thus its inverse Fourier
transform will produce a map in units of Jy/beam, where the beam area is 
given by the point-spread-function (PSF) of the image.  For a single field,
the PSF is just the image generated using equation (\ref{eq:image}) using
estimators $\Delta^{\rm PSF}_i$ computed by introducing unit ``visibilities'' 
into equation (\ref{eq:linop})
\begin{equation} \label{eq:psf}
  \Delta^{\rm PSF}_i = \sum_k \big\{ Q_{ik} + \overline{Q}_{ik} \big\}.
\end{equation}
The situation for the mosaics is somewhat more complicated, as the
mosaic offsets must be taken into account in constructing equivalent
visibilities for point sources
\begin{eqnarray} \label{eq:vpsf}
  \Delta^{\rm PSF}_i(\hat{\bvec{x}}) & = & 
     \sum_k \big\{ Q_{ik}\,V^{\rm PSF}_k(\hat{\bvec{x}}) 
     + \overline{Q}_{ik}\,V^{\ast\rm PSF}_k(\hat{\bvec{x}}) \big\} 
     \nonumber \\
  V^{\rm PSF}_k(\hat{\bvec{x}}) & = & A_{k}(\hat{\bvec{x}}-\bvec{x}_k)
     e^{-2\pi i \sbvec{u}_k\cdot(\hat{\sbvec{x}}-\sbvec{x}_k)}
\end{eqnarray}
obtained by evaluating equation (\ref{eq:visi}) with no noise and
$I(\bvec{x}) = \delta^2(\bvec{x}-\hat{\bvec{x}})$.  
In this case one would evaluate
the PSF at various positions $\hat{\bvec{x}}$ in the map.

Because our estimators use the kernel ${\bf Q}$ as given in equation
(\ref{eq:qik}) which includes the beam transform $\tilde{A}$, we are
effectively multiplying the image on the sky by the primary beam squared
--- once in the kernel, and once due to the instrument itself 
(e.g.\ eq.[\ref{eq:pk}]).  Images made directly from the $\bvec{\Delta}$ will
therefore be strongly attenuated in the (noisy) outskirts.

As mentioned in Appendix~\ref{app:estim}, the optimal weighting
for the imaging of the CMB component is to use the Planck factor in
equation (\ref{eq:fplanck})
to correct for the thermal frequency spectrum (e.g.\ eq.[\ref{eq:qtherm}]),
while our standard intensity weighting given in equation (\ref{eq:viswt})
is optimized for a flat non-thermal power-law spectrum with spectral index
$\alpha=0$.

We can also filter the gridded estimators in such a way as to enhance
or down-weight certain signals or noise.  We can do this with optimal
or Wiener filtering (e.g.\ \citealt{bc01}), 
\begin{equation}
    \bvec{\Delta}^\phi = {\bf \Phi}\bvec{\Delta} 
\end{equation} 
where the choice of the filter ${\bf \Phi}$ depends on the application.
For example, the covariances ${\bf C}^{\rm X}$ calculated from
equation (\ref{eq:csbreimest}) can be used to construct optimal filters 
for each component
contributing to the observations. For the signal component
described by the covariances ${\bf C}^{\rm X}$ we can construct a Wiener
filter to be applied to the gridded $uv$ estimators
\begin{equation}
   \bvec{\Delta}^{\rm X} = {\bf C}^{\rm X}\,{\bf C}^{-1}\,\bvec{\Delta}.
\end{equation} 
The amplitudes for the signal models such as the bandpowers $q_B$ or the
source amplitudes $q_{\rm src}$ can be set to their Maximum Likelihood
values or to fiducial model amplitudes. The Wiener
filtered image is then recovered by Fourier transforming.

Examples of images created using this method are described in the next
section, and shown in Figure~\ref{fig:image}.  Wiener-filtered images
are also used in Paper~VI to explore the possibility of detection of
the Sunyaev-Zeldovich effect in the data at high $\ell$.

\section{Implementing the Method} \label{sec:estliklhd}

The algorithm described above was coded as a scalar {\sc Fortran}
({\tt f77} compatible) program designated {\tt cbigridr}, with a parallelized
{\sc Fortran 90} version 
using OpenMP\footnote{\tt http://www.openmp.org/} directives
also available for use on multiprocessor machines.  In addition,
the BJK likelihood relaxation was coded in a second parallelized {\sc Fortran
90} program called {\tt mlikely} using parallel versions of
{LAPACK} matrix algebra routines (e.g.\ \S\ref{sec:relax}).  
Together, these two programs make up the CBI analysis pipeline.  
This pipeline has undergone numerous tests and
development since its inception in April 2001, and has been used to 
produce the power spectra and to provide the bandpowers as input to the
cosmological parameter analysis given in the companion papers.  We now
give a brief description of our implementation.

In order to carry out the numerical integrations, a fine-grain rectangular
lattice in {\it uv}-space was used.  The fine-grain grid size 
$\Delta u_{\rm fine}$ (in units of wavelength) was chosen to adequately sample
the phase turns in the 
{\it uv} plane due to the mosaic size and differencing; for a standard
$7\times6$ CBI mosaic with $20^\prime$ spacing, the maximum field separation
along the grid direction is $x_{\rm max}= 2^\circ$, which gives 
oscillations in the {\it uv} plane with wavelength
of $x_{\rm max}^{-1} = 28.65$, giving $\Delta u_{\rm fine} \leq 14.3$ for two
samples per cycle.  To evaluate the projection operators $R_i(\bvec{v})$ 
(e.g.\ eq.[\ref{eq:rop}]), we store a small fine-grain lattice around each 
$\bvec{u}_i$.  The maximum radius in {\it uv} space needed for the support of
${\bf R}$ is $r_u=2D/\lambda_{\rm min}$.  For CBI $D=90$~cm, and
$\lambda_{\rm min}=0.844$~cm at $35.5$~GHz, we get $r_u=213.1$.  
A grid of size $53\times53$ cells with $\Delta u_{\rm fine}=8.526$
will fit both the sampling and radius requirements.

The estimators are evaluated on a coarse-grain lattice of $\bvec{u}_i$,
with a spacing of $\Delta u_{\rm coarse}$.  The fine-grain lattices
on which we accumulate the $R_i(\bvec{v})$ will be cross-correlated to
form the covariance elements ${\bf M}_B$ and
$\overline{\bf M}_B$ (e.g.\ eq.[\ref{eq:mkkint}]), it is desirable to have the
coarse grid size locked to integer multiples of the fine-grid cells.  This
coarse grid  does not have to sample the highest mosaic frequencies, but only
the effective width of ${\bf R}$.  We find that for single CBI fields,
$\Delta u_{\rm coarse}=3\,\Delta u_{\rm fine}$ is adequate.  For CBI mosaics,
a hybrid lattice with $\Delta u_{\rm coarse}=\Delta u_{\rm fine}$ in the inner
part ($\ell<800$) and $\Delta u_{\rm coarse}=2\,\Delta u_{\rm fine}$ 
in the outer part was found to work well.

As stated in \S\ref{sec:pscmb}, the choice of sign of the exponential
of the Fourier transform in equation (\ref{eq:fourier}) is a convention.  This
choice varies throughout the literature on the
subject, but in practice depends upon the way the baseline vectors
are defined in the data and how the correlation products are made
(e.g.\ which antenna gets the quadrature phase shift).  
We note that in coding our algorithm to conform to the imaging
standards of the AIPS\footnote{\tt http://www.cv.nrao.edu/aips/}, and 
DIFMAP\footnote{\tt ftp://ftp.astro.caltech.edu/pub/difmap/} \citep{difmap} 
packages using the CBI data, we had to use the opposite
sign convention from the one presented in equation (\ref{eq:fourier}).

To process a dataset, a spectral weighting $f(\nu)$ and shape function
$C^{\rm shape}_\ell$ are chosen.  The visibilities $V_k$ are looped over, 
and any source subtraction (\S\ref{sec:src}) is applied.
For each estimator $i$ that $V_k$ contributes to either directly or as a
conjugate, 
its contribution to the fine-grain lattice $q$ for $R_i(\bvec{v}_q)$
is accumulated, e.g.
\begin{equation}
   R_i(\bvec{v}_q) = \sum_k \{ Q_{ik}\,P_k(\bvec{v}_q) + 
    \overline{Q}_{ik}\,\overline{P}_k(\bvec{v}_q) \}
   \hspace{2cm} \bvec{v}_q = \bvec{u}_i + \Delta u_{\rm fine}\,\hat{\bvec{v}}_q
\end{equation}
where $\hat{\bvec{v}}_q$ is a $53\times53$ unit (fine-grain) lattice.
This means that for $n_{\rm est}$ estimators, the storage required for 
${\bf R}$ is only $2809\times n_{\rm est}$ double-precision complex numbers.
If the data were differenced (as for CBI data), then $P^{\rm sw}_k(\bvec{v})$
from equation (\ref{eq:pksw}) is used.
The contributions to the noise covariance elements ${\bf M}^{\rm N}$ and 
$\overline{\bf M}^{\rm N}$ (eq.[\ref{eq:eeexmat}]), and the
$\bvec{\Delta}^{\rm src}_{c}$  (eq.[\ref{eq:dsrcrms}]) are also accumulated at
this time.  Finally, this visibility's contributions are added to
estimator $\Delta_i$ and normalization $z_i$.

The storage for ${\bf R}$, ${\bf M}^{\rm N}$, and $\overline{\bf M}^{\rm N}$
dominate the memory requirements. 
For example, the CBI mosaics use around $n_{\rm est}=2500$ estimators, and thus
storage for $53\times53\times2500\approx7\times10^6$ double-precision
complex numbers is needed.  A single packed $n^2_{\rm est}\approx6\times10^6$
array is needed to hold ${\bf M}^{\rm N}$ and $\overline{\bf M}^{\rm N}$.  
The ${\bf C}^{\rm X}$ matrices
are calculated in place and written out row by row, and thus need not be
stored.  There are no instances where matrices of dimension $N_{\rm vis}^2$ are
stored; the storage for a matrix of this size would be prohibitive as our
largest CBI mosaics have $N_{\rm vis}>2\times10^5$. 

When all the visibilities have been processed, the estimators are normalized
by $z_i$, split into real and imaginary parts, and written out to disk.
The covariance matrices $C^{\rm N}_{ij}$, $C^{\rm S}_{Bij}$, and any 
$C^{\rm src}_{ij}$  
are constructed by looping over pairs of rows corresponding to the real
and imaginary parts of each estimator, e.g.\ rows $i$ and $i+N_{\rm est}$ for
estimator $i$.  For each $j\leq i$, the stored $R_i$ and $R_j$ are 
cross-correlated
along with the shape function ${\cal C}(|\bvec{v}|)$ to form the bandpower
covariance elements $M_{Bij}$ and $\overline{M}_{Bij}$ of 
equation (\ref{eq:mkkint}), combined to make $C^{\rm S}_{Bij}$ using equation
(\ref{eq:csbreimest}), and stored.  The relevant columns of these rows
of $C^{\rm N}_{ij}$ are formed from the stored $M^{\rm N}_{ij}$ and
$\overline{M}^{\rm N}_{ij}$. 
At this point, for each ${\bf C}^{\rm src}$ desired (there may be more than
one, in the CBI analysis we use 3), the relevant $\Delta^{\rm src}_c$ are
combined using equation (\ref{eq:mkksrc}) to form $M^{\rm src}_{ij}$ and
$\overline{M}^{\rm N}_{ij}$, which in turn are used 
to make $C^{\rm src}_{ij}$.  After all columns for these rows of the covariance
matrices are computed, they are written to disk, and this process is repeated
for the pair of rows corresponding to the next estimator $i$.  When all
rows are complete, the output file is complete.  Note that 
different binnings of the ${\bf C}^{\rm S}_{B}$ can be run without regridding
using the original ${\bf R}$, saving significant time.

If a residual foreground covariance matrix ${\bf C}^{\rm res}$ is desired,
the procedure outlined above is repeated in its entirety using the 
description in \S\ref{sec:foreground}.  The spectrum and shape appropriate
for the source population or foreground emission is used during the
gridding and covariance matrix construction.  Other than these factors, the
same gridding as in the CMB and noise estimators must be used.

The output files from {\tt cbigridr} are then used as input to {\tt mlikely}.
These can be for single fields or mosaics, or for combinations of independent
fields or mosaics (\S\ref{sec:comblike}).
At this point, the pre-factors $q_{\rm src}$ and $q_{\rm res}$ for any ${\bf
C}^{\rm src}$ and ${\bf C}^{\rm res}$ covariance matrices are chosen and
fixed.  Relaxation to the likelihood maximum is carried out as described in
\S\ref{sec:relax}, 
and the resulting bandpowers $\{q_B\}$ and inverse Fisher matrix elements
$[ F^{-1} ]_{B B'}$ are written out.  If desired, the bandpower window
functions $W^B_{B_f}$ (\S\ref{sec:bwin}) can be computed if {\tt cbigridr} was
run to produce narrow-bin ${\bf C}^{\rm S}_{B_f}$.  The component bandpowers
$q_B^{\rm N}$, $q_B^{\rm src}$ and $q_B^{\rm res}$ (\S\ref{sec:comp}) can also be
computed at this time.

Finally, filtered images using the formalism of \S\ref{sec:image} can 
be computed from the estimators, the ${\bf C}$ (at maximum likelihood),
and the component covariance matrices.  Results from this are shown below and
in Paper~VI.

The timing for {\tt cbigridr} depends upon the degree of parallelization as
well as the processor speed on a given machine, and the number of visibilities
gridded, number of foreground sources, and number of bins $B$ for
the bandpowers.  As an example, the 
processing of the 14-h mosaic field of Paper~III (the largest of the
datasets) involved gridding 228819 visibilities from 65 separate
nights of data in 41 fields to 2352 complex estimators.
A total of 916 sources were gridded into three source covariance matrices.
A total of 7 different binnings for ${\bf C}^{\rm S}_{B}$ were run at this
time from the same gridding.  The execution time using
the parallel version of {\tt cbigridr} was
$2^{\rm h}\,40^{\rm m}$ running on 22 processors on a 32-processor 
Alpha GS320 workstation at the Canadian Institute for Theoretical Astrophysics.
It then took $3^{\rm h}\,22^{\rm m}$ on the same computer for {\tt
mlikely} to process 4704 double-precision real estimators in 
16 ${\bf C}^{\rm S}_{B}$ bands, with 3 ${\bf C}^{\rm src}$ matrices,
one ${\bf C}^{\rm res}$ and one ${\bf C}^{\rm N}$.  
This included the time needed to calculate the component bandpowers 
${\bf C}^{\rm X}_{B}$, but not the window functions.  
The speed of this fast gridded method has allowed us to carry out
numerous tests on both real and simulated dataset, which would not have been
possible carrying out maximum likelihood (e.g.\ using even the
optimized {\tt mlikely}) on the 200000-plus visibilities.

\subsection{Method Performance Tests Using Mock CBI Data} \label{sec:test}

The performance of the method was assessed by applying it to mock CBI
datasets.
Simulated CBI datasets were obtained by replacing the actual visibilities
from the data files containing real CBI observations of the various
fields used in Paper~II and Paper~III with the response expected for 
a realization of the CMB sky drawn from a representative power spectrum, plus
uncorrelated Gaussian instrumental noise with the same variance as given by
the scatter in the actual CBI visibilities. 
The differencing of the lead and trail fields used in CBI observations was
included (e.g.\ \S\ref{sec:difference}).
This mock dataset had the same {\it uv} distribution as the
real data, and gives an accurate demonstration of expected
sensitivity levels and the effect of cosmic variance. The power spectrum
chosen for these simulations was for a model that fit the COBE and
BOOMERANG data \citep{ne01}.

\begin{figure}
\includegraphics[angle=-90,width=6.5in]{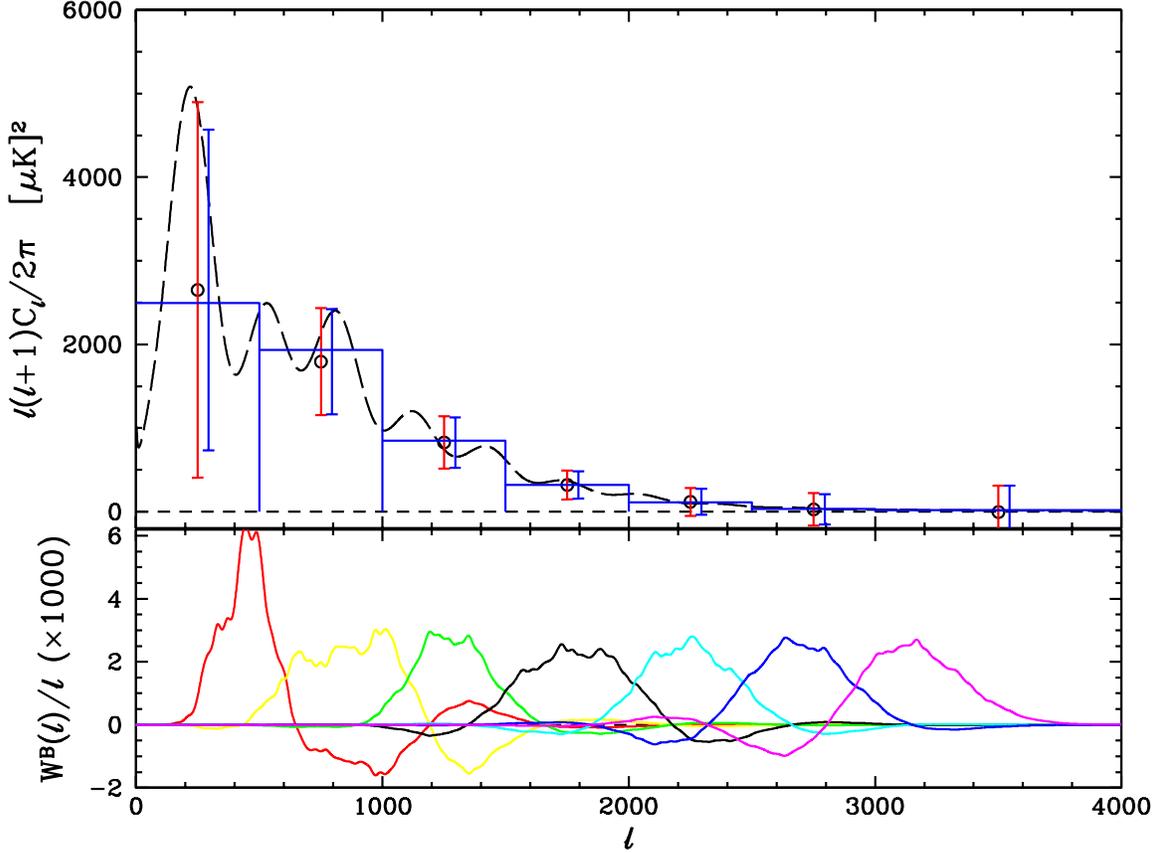}
\caption{Results of the gridded method plus quadratic
relaxation for 387 mock CBI 08h deep-field datasets (upper panel), 
with each mock observation drawn from an independent realization of the
sky given the model power spectrum (shown as the dashed curve) 
and the instrumental noise with the appropriate rms.  
The points (black circles) are placed at the
band centers, at the mean of the reconstructed bandpowers with the (red) 
errorbars given by the scatter among the realizations.  The (blue) errorbars
to the right of the points show the average of the inverse Fisher matrix
diagonals.  The 
histograms show the width of each band and the level expected by integrating
the model $\cal{C}_\ell$ over the window functions $W^B_{\ell}$ which are
shown in the lower panel.  The mean of the realizations converges to the
expected value within the Poisson uncertainty, taking into account the
correlations between adjacent bands.\label{fig:mockdeep}}
\end{figure}

\begin{figure}
\includegraphics[angle=-90,width=6.5in]{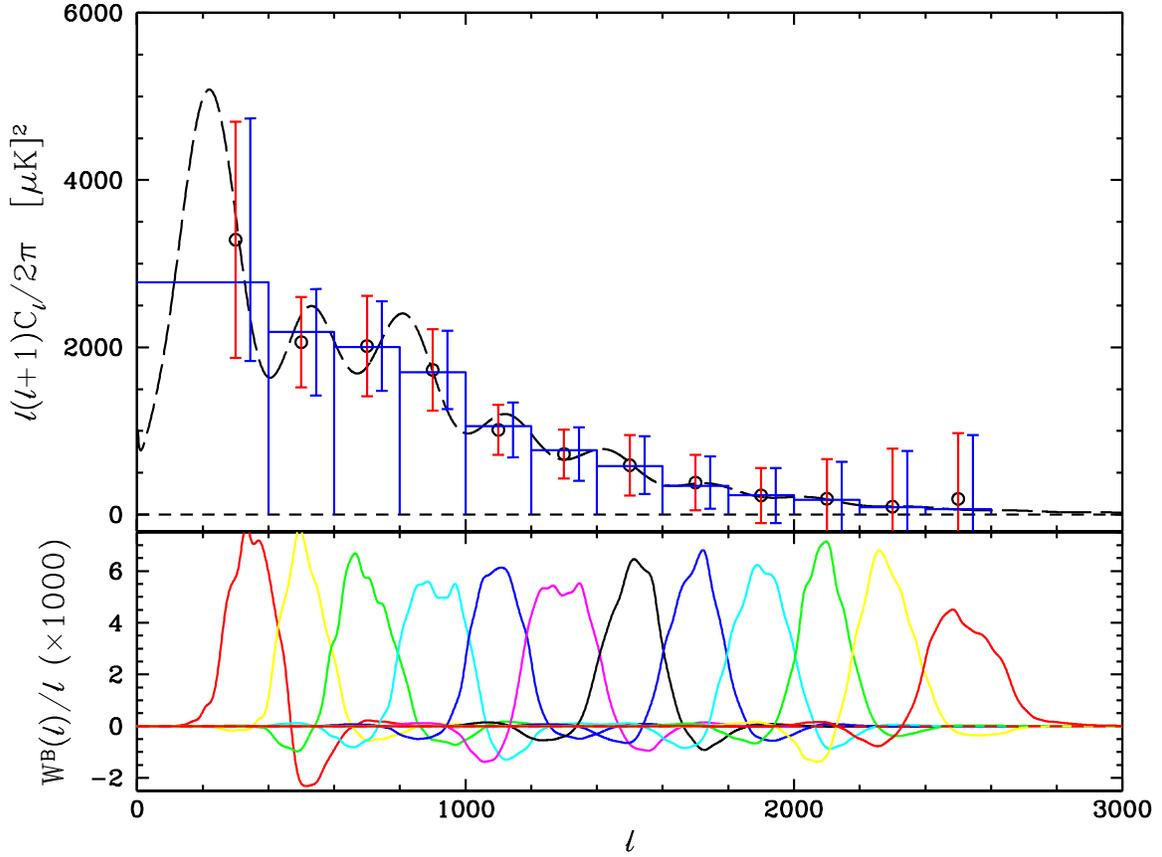}
\caption{Upper panel shows the results of the gridded method plus
quadratic relaxation for 117 mock CBI ($7\times6$ field) mosaic datasets 
with a CDM-based power spectrum (the dashed curve).  
As in Figure~\ref{fig:mockdeep}, the 
points (black circles) are centered in each bin in $\ell$ with (red) errorbars
giving rms scatter of about the mean for the bandpowers from the processed
realizations, with the (blue) error bar from average inverse Fisher diagonals 
to the right of the points.  The window functions are plotted in the
lower panel.\label{fig:mockmos}}
\end{figure}

Figure~\ref{fig:mockdeep} shows the power spectrum estimation derived
following the procedure detailed above.  The mock datasets were drawn 
as realizations for the 08h CBI deep field from Paper~II. The binning of the
signal covariance matrix ${\bf C}^{\rm S}_{B}$ was chosen to be uniform in
$\ell$ with bin width $\Delta\ell=500$.
Because a single realization of the sky drawn from the model power spectrum
will have individual mode powers that deviate from mean given by the power
spectrum due to this intrinsic so-called ``cosmic variance'' plus the
effect of the thermal instrumental noise, we analyze 387 realizations
each taken from a different realization of the sky and a different set
of instrumental noise deviates.  The mean $q_B$ for each band $B$ converge
to $\langle C_B\rangle$, which is obtained by
integrating the model $\cal{C}_\ell$ over the window functions $W^B_{\ell}$
(e.g.\ eq.[\ref{eq:cbpred}]), within the sample uncertainty for the
realizations.  Furthermore,  the standard deviation of the $q_B$ from the mean
for each band agrees with the value obtained from the diagonals of the inverse
of the Fisher matrix.

The choice of the $\ell$ bin size is driven by the trade off between
the desired narrow bands for localizing features in the power spectrum and
the correlations between bins introduced by the transform of the primary
beam.  There is an anti-correlation between adjacent bands seen
in $[F^{-1}]_{BB'}$ at the level of $-13$\% to $-23$\% for $\Delta\ell=500$
with a single field. We have found that correlations up to about $-25$\%
give plots of the $q_B$ that are more visually appealing than those made
with narrower band and higher correlation levels due to the increasing
scatter in the bandpowers about the mean values.  Bins of this
size do not achieve the best possible $\ell$-resolution, and thus our
cosmological parameter runs use finer-binned bands since the correlations
are taken into account in the analyses.

The band window functions $W^B_\ell$ are shown in the lower panel
of Figure~\ref{fig:mockdeep}, and were computed using narrow binnings
$W^B_{B_f}$ (e.g.\ eq.[\ref{eq:wbbf}]) with $\Delta\ell=20$.
The small-scale structures seen in the window functions, particularly visible
around the peaks, are due to the differencing which introduces oscillations
(see \S\ref{sec:difference}).
As shown in equation (\ref{eq:winortho}), a window function $W^B_\ell$
is normalized to sum to unity within the given band $B$, and to
sum to zero in the other bands, and thus there must be compensatory
positive and negative ``sidelobes'' of the window function outside the band.

Figure~\ref{fig:mockmos} shows the power spectrum derived for a simulated
mosaic of $7\times6$ fields separated by $20^\prime$ using the actual
CBI 20h mosaics fields from Paper~III as a template.  This mosaic field
was chosen as it had incomplete mosaic coverage, and thus would be the
most difficult test for the method.  The binning for ${\bf C}^{\rm S}_{B}$
shown used $\Delta\ell=200$, which gave adjacent band anti-correlations of
$-13\%$ to $-18\%$ in the $F_{BB'}$.  Again the mean of the 117
realizations converges to the value expected within the error bars, showing
that there is no bias introduced by the method, even in the presence
of substantial holes in the mosaic (see Paper~III for the mosaic weight
map).  Furthermore, the rms scatter in the realizations converges to
the mean of the inverse Fisher error bars, as in the single-field case. 
As in the previous figures, the bandpower window
functions are shown in the lower panels.  

\begin{figure}
\includegraphics[angle=-90,width=6.5in]{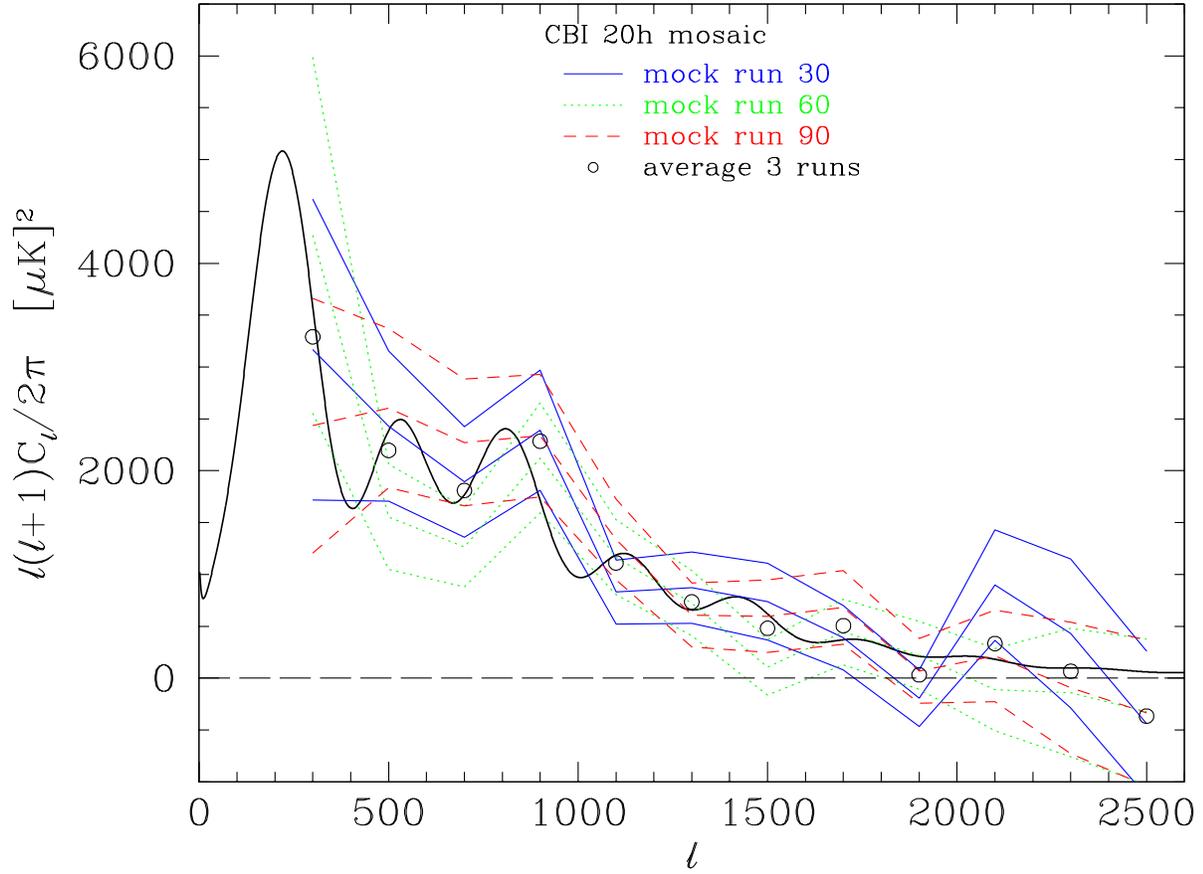}
\caption{Three randomly selected realizations from the mosaic field
simulations shown in Figure~\ref{fig:mockmos} are shown plotted against
the model power spectrum (solid black curve).  The three lines for
the bandpowers reconstructed from the three realizations correspond to
the bandpowers (central lines) and the $\pm1\,\sigma$ excursions using
the inverse Fisher errorbars.  The scatter in bandpowers between realizations
is within the expected range.  Also shown is the unweighted scalar average
of the bandpowers for the 3 realizations, which is an approximation to
a true joint likelihood solution.  The average is a better fit to the
model, as is expected.\label{fig:mockmosruns}}
\end{figure}

In Figure~\ref{fig:mockmosruns} 
are shown three randomly chosen realizations from the ensemble, plotted 
along with the input power spectrum.   This shows the level of field-to-field
variations that we might expect to see in CBI data.
There are noticeable deviations from the expected bandpowers in individual
realizations, particularly at low $\ell$ where cosmic variance and the
highly-correlated bins conspire to increase the scatter.  These differences
are within the expected scatter when bin-bin correlations and limited sample
size is taken into account, but care must be exercised in interpreting
single field power spectra.  In particular, the acoustic peak structures are
obscured by the sample variations.  However, the average bandpowers for the 3
runs (shown in Figure~\ref{fig:mockmosruns} as open black circles) are better
representations of the underlying power spectrum.  Although this is not a
proper ``joint'' maximum likelihood solution (e.g.\ \S\ref{sec:comblike})
as is done for the real CBI mosaic fields, the improvement seen using
the 3-field average leads us to expect that the combination of even
3 mosaic fields damps the single field variations sufficiently
to begin to see the oscillatory features in the CMB power spectrum.  While
we do not show the equivalent plots of the deep fields from 
Figure~\ref{fig:mockdeep}, the same behavior is seen (with even larger
field-to-field fluctuations in the relatively unconstrained first bin,
though still consistent with the error bars).

\begin{figure}
\includegraphics[angle=-90,width=6.5in]{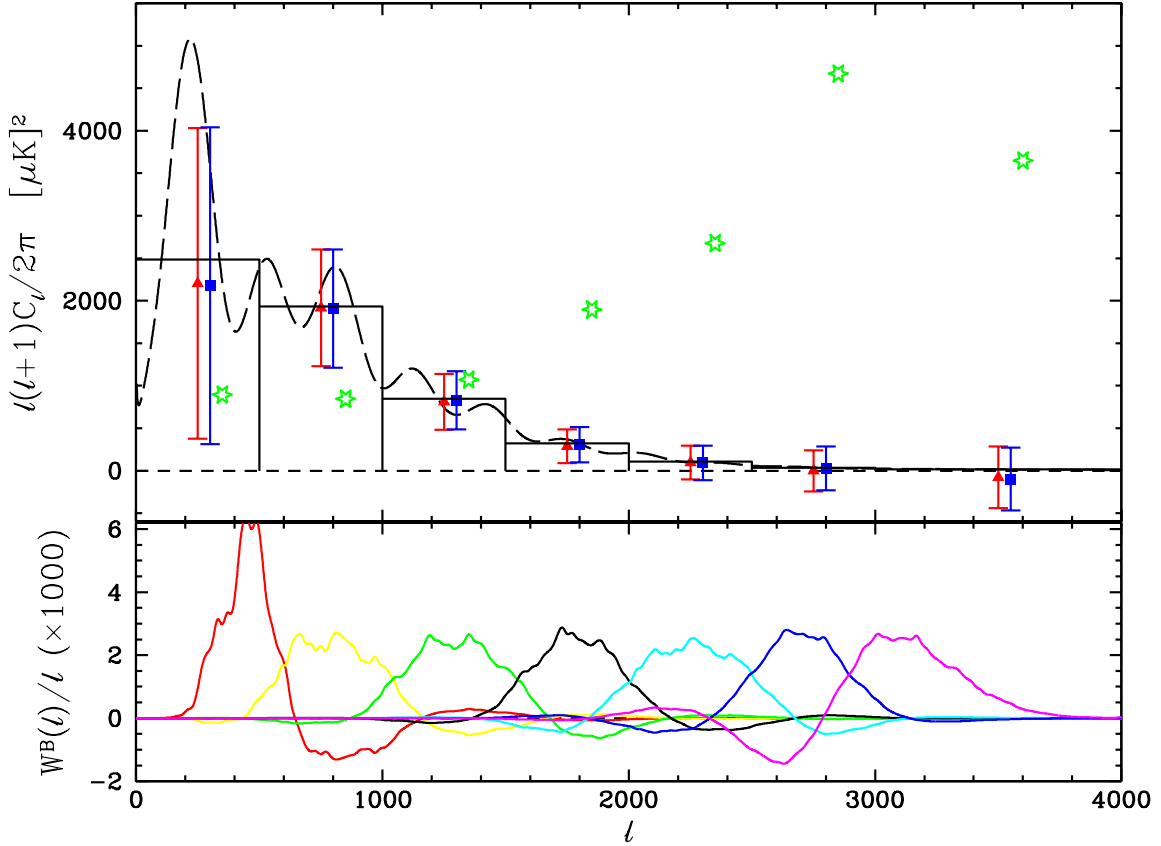}
\caption{Results using mock deep-field datasets
including foreground point sources based on the actual list used in the CBI
data are shown (upper panel) along with the window functions (lower panel).
The input power spectrum and expected bandpowers are as in 
Figure~\ref{fig:mockdeep}.  The (green) stars show the average for 200
realizations where no source 
subtraction or projection was done, with the powers divided by a factor of 2
to fit on the plot. The expected increase with $\ell^2$ is seen, along with a
falloff in the last bin due to the source frequency spectrum.  The points with
errorbars at the center of each bin (red triangles) were computed from 200
realizations processed with subtraction of the mean flux density from
the visibilities and construction of ${\bf C}^{\rm src}$ using the
uncertainties, while the points with errorbars to the right of these (blue
squares) are from 200 realizations where no source subtraction was done, but
we built ${\bf C}^{\rm src}$ using the full (unperturbed) flux densities which 
projects out the sources with only a slight increase in noise.  Despite
the large power from sources at high $\ell$, our method successfully 
removes the foreground power from the spectrum with no sign of bias.
\label{fig:mockdeepsrc}}
\end{figure}

\begin{figure}
\begin{tabular}{cc}
\makebox[3in][c]{\includegraphics[width=2.5in]{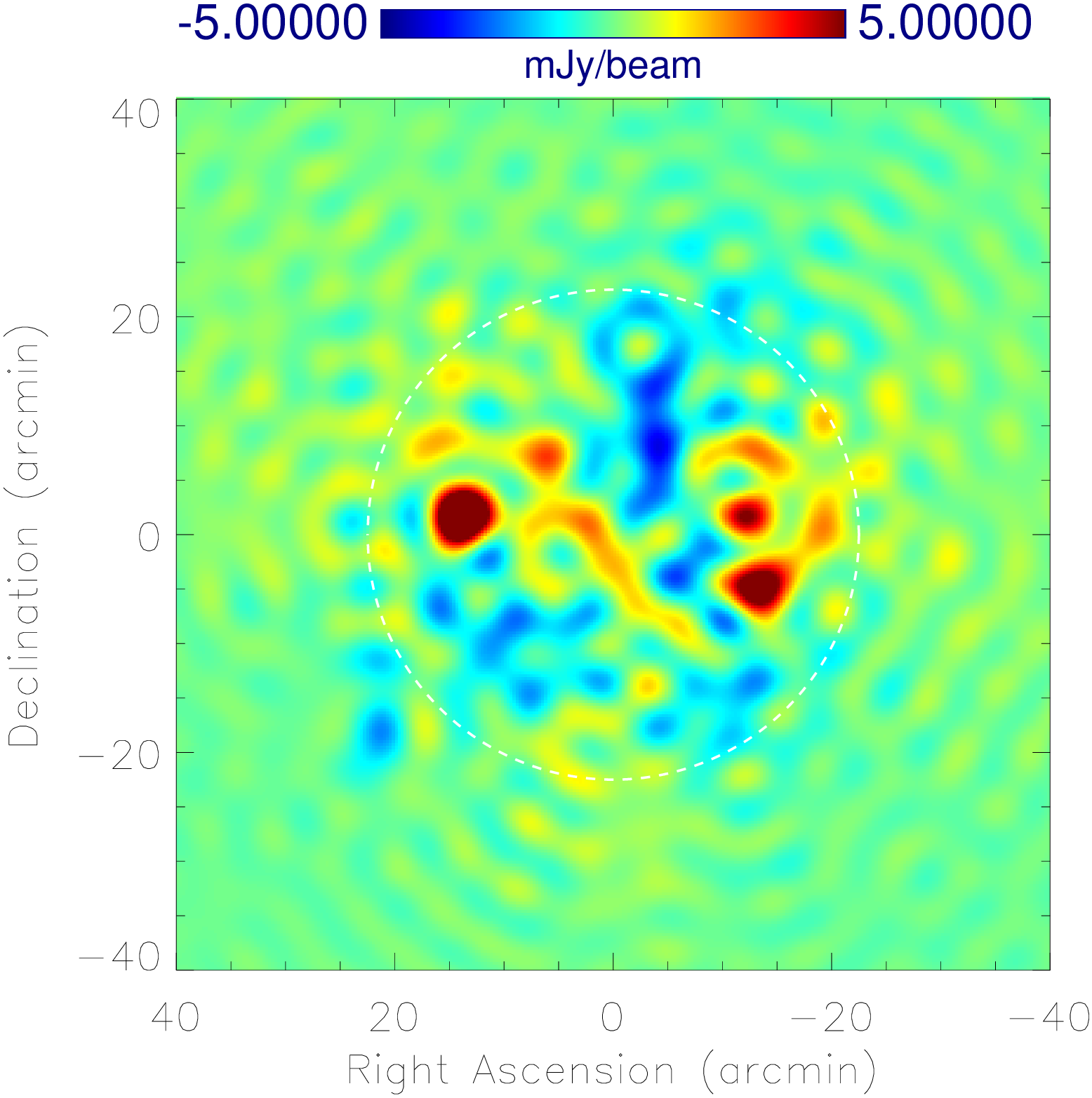}} &
\makebox[3in][c]{\includegraphics[width=2.5in]{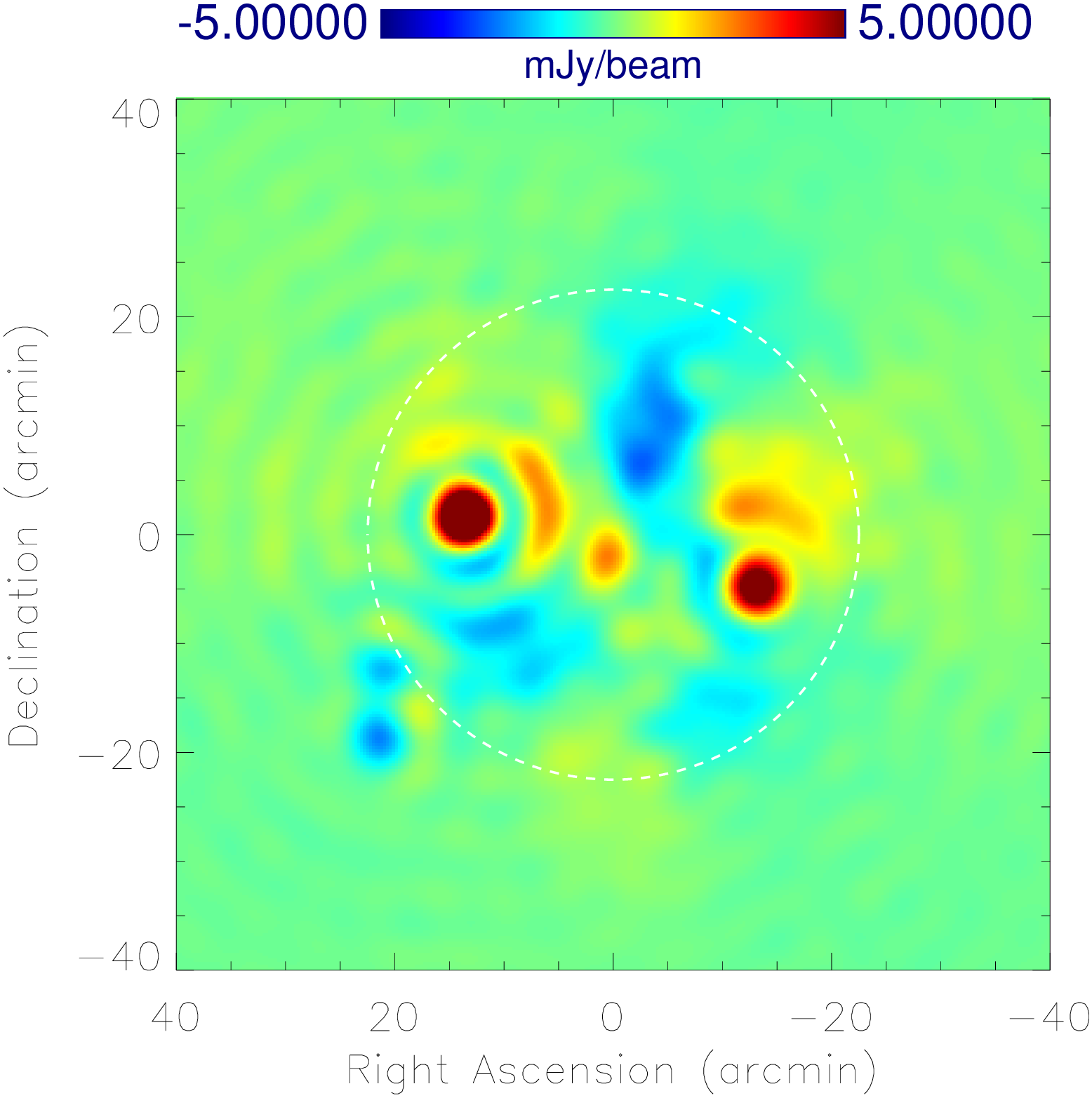}} \\[1cm]
\makebox[3in][c]{\includegraphics[width=2.5in]{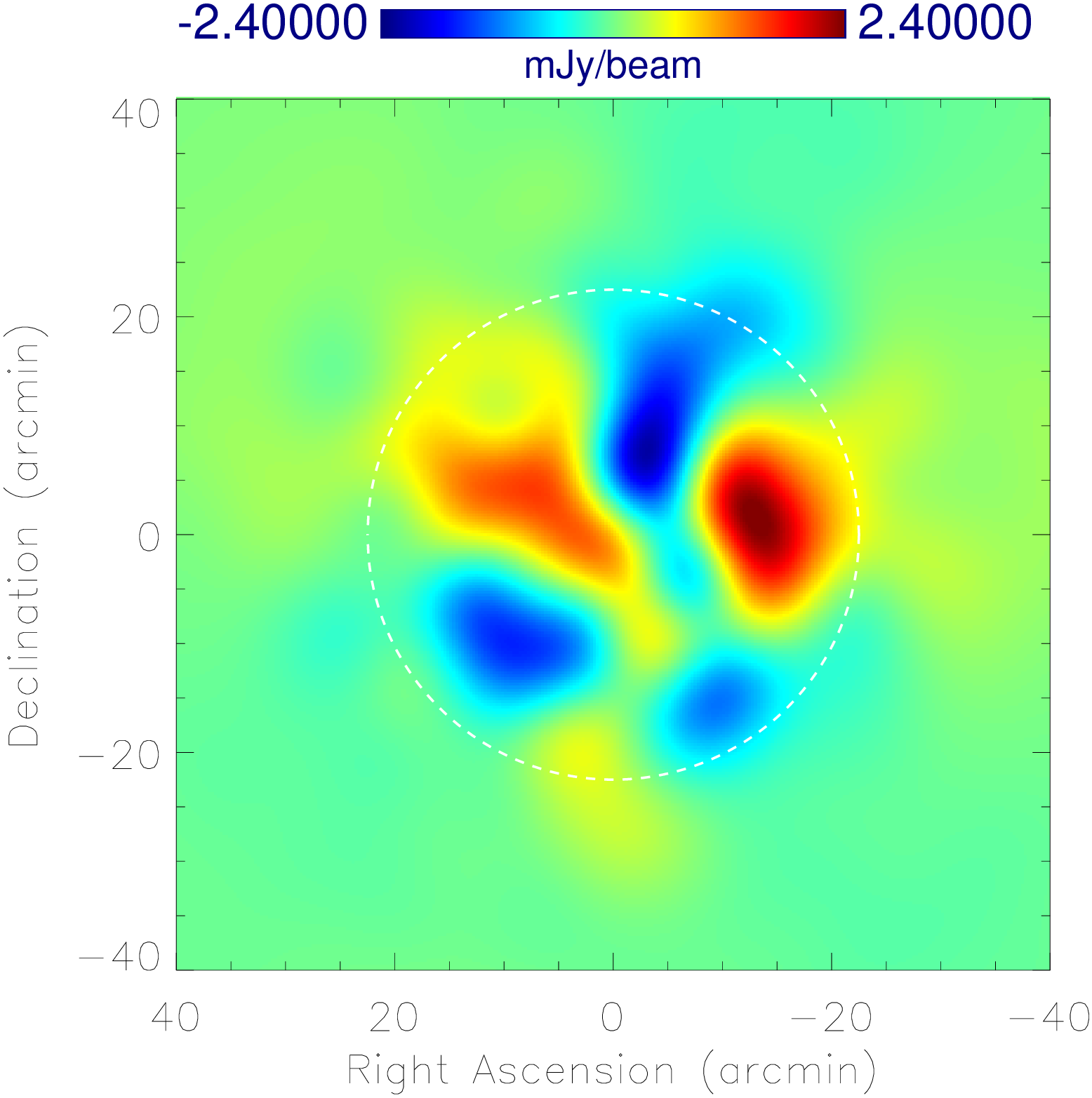}} &
\makebox[3in][c]{\includegraphics[width=2.5in]{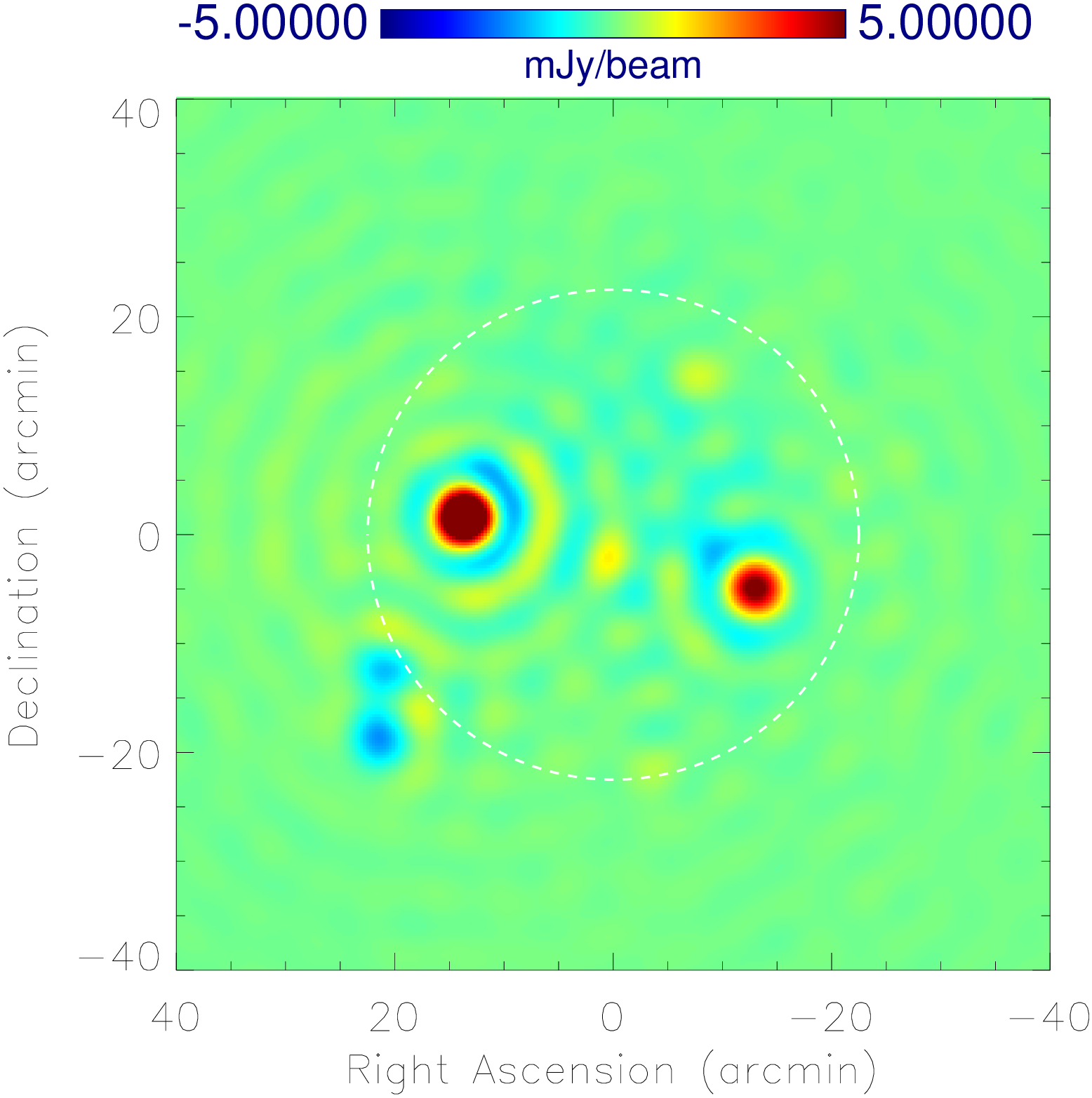}} \\[1cm]
\end{tabular}
\caption{ Images reconstructed from the gridded estimators using the
formalism of \S\ref{sec:image}.  Data is for one of the mock 08h deep-field
realizations with sources used in Figure~\ref{fig:mockdeepsrc}.
Shown are: Panel (a) upper left --- image computed without any filtering;
(b) upper right --- the image derived by setting 
${\bf C}^{\rm X} = {\bf C}^{\rm src} + \sum_B q_B\,{\bf C}^{\rm S}_B$ 
the sum of the signal terms; (c) lower left --- the image using 
${\bf C}^{\rm X} = \sum_B q_B\,{\bf C}^{\rm S}_B$ for the CMB only;
(d) lower right --- image for ${\bf C}^{\rm X} = {\bf C}^{\rm src}$ to
pick out the point sources.  The filter clearly dampens the
noise, and separates the CMB and source components.  The residuals from
several bright sources dominate the signal in all but the CMB-filtered image
(the brightness scale in that image is enhanced).  The white dashed circle
shows the $45\farcm2$ FWHM of the CBI at 31~GHz.  The attenuation of the
signal brightness due to the square of the primary beam is clearly seen.  
\label{fig:image} }
\end{figure}

The effect of adding point sources to the mock fields, and then attempting
power spectrum extraction, is shown in Figure~\ref{fig:mockdeepsrc}.  A
set of 200 realizations were made in the same manner as in the runs in
Figure~\ref{fig:mockdeep}, 
but the list of point source positions, flux densities and uncertainties,
and spectral indices from lower frequency used in
the analysis in Paper~II (the ``NVSS'' sources) was used to add mock sources
to the data. The flux density of the sources actually added to the data were
perturbed using the stated uncertainties as 1-$\sigma$ standard
deviations.  The errors used were 33\% of the flux density except for a few of
the brighter sources which were put in with 100\% uncertainties.  
We then used the methodology described in \S\ref{sec:src} to compute
the constraint matrices.  The first method of correction
used was to subtract the (unperturbed) flux densities from the visibilities,
and build the ${\bf C}^{\rm src}$ from $\bvec{\Delta}^{\rm src}$ built using
the uncertainties (shown as the red triangles).  In addition, we also 
did no subtraction, but built ${\bf C}^{\rm src}$ from 
$\bvec{\Delta}^{\rm src}$  
using the full (unperturbed) flux densities (shown as the blue squares).
This is equivalent to assuming a 100\% error on the source flux densities, and
thus canceling the average source power in those modes.  In both cases
the factor $q_{\rm src}=1$ was used.  The simulations show that both methods
are effective, with no discernible bias in the reconstructed CMB bandpowers.

Finally, the production of images using the gridded estimators described
in \S\ref{sec:image} is demonstrated in Figure~\ref{fig:image}.  
The series of plots show the effect of Wiener filtering using the noise and
various signal components on an image derived from one of the mock 08h CBI
deep field realizations with sources from the ensemble shown in 
Figure~\ref{fig:mockdeepsrc}.  The Planck factor weighting of equation
(\ref{eq:qtherm}) was used during gridding to optimize for the thermal CMB
spectrum, though in practice this makes little difference due to the
restricted frequency range of the CBI.
The estimators for this realization were
computed by subtracting the mean values of the source flux densities and
putting the standard deviations into ${\bf C}^{\rm src}$ with $q_{\rm src}=1$
(the red triangles in Figure~\ref{fig:mockdeepsrc}).
The filtering down-weights the
high spatial frequency noise seen in the unfiltered image, and effectively 
separates the CMB and source components as shown by comparing panels (c) 
and (d) to the total signal in panel (b).  The signal in this realization
is dominated by the residuals from two bright point sources that had 100\%
uncertainties put in for their flux densities and thus escaped subtraction.
The effectiveness of ${\bf C}^{\rm src}$ in picking out the sources in the
image plane underlines its utility as a constraint matrix in the power spectrum
estimation.

\section{Conclusions}\label{sec:conclude}

We have outlined a maximum likelihood approach to determining the
power spectrum of fluctuations from interferometric CMB data.  
This fast gridded method is able to handle the large amounts
of data produced in large mosaics such as those observed by the CBI.  
Software encoding this algorithm was written, and tested using mock CBI
data drawn from a realistic power spectrum.  The results of
the code were shown to converge as expected to the input power spectrum
with no discernible bias.  For small datasets, this code was also tested
against independently written software that worked directly on the 
visibilities.  In addition, the pipeline was run with gridding turned
off as described in \S\ref{sec:liklhd}, again for small test data sets. 
No bias or significant loss in sensitivity was seen in these comparisons.  

This software pipeline was used to analyze the actual CBI data, producing the
power spectra presented for the deep fields and mosaics in Paper~II and
Paper~III respectively.  The output of the pipeline also was used as the input
for the cosmological parameter analysis in Paper~V and the investigation
of the Sunyaev-Zeldovich Effect in Paper~VI.

This method is of interest for carrying out power spectrum estimation
for interferometer experiments that produce
a large number of visibilities but with a significantly smaller number
of independent samples of the Fourier plane (such as close-packed arrays
such as VSA or DASI).  The CBI pipeline analysis is carried out in
two parts, the gridding and covariance matrix construction from input
uv-FITS files in {\tt cbigridr} and the maximum likelihood estimation
of bandpowers using quadratic relaxation in {\tt mlikely}.  The software
for the pipeline is available by contacting the authors.

We close by noting that our formalism can be extended to deal with
polarization data.  In the case of CMB polarization, there are as many
as six different signal covariance matrices of interest in each band,
with estimators (or visibilities) for parallel-hand and cross-hand polarization
products, and thus development of a fast method such as this is critical.  In 
September 2001 polarization capable versions of {\tt cbigridr} and {\tt
mlikely} were written and tested.  We describe the method, the polarization
pipeline, and results in the upcoming paper (Myers et al.\ 2002, in
preparation).

\acknowledgments

STM was supported during the early years of the CBI by a Alfred P.\ Sloan
Fellowship from 1996 to 1999 while at the University of Pennsylvania.  Genesis
of this method by STM greatly benefited by a stay in July 2000 at the ITP in
Santa Barbara, supported in part by the National Science Foundation under
grant PHY99-07949.  The National Radio Astronomy Observatory
is a facility of the National Science Foundation operated under cooperative
agreement by Associated Universities, Inc.  The CBI was funded under 
NSF grants AST-9413934, AST-9802989 and AST-0098734, with 
contributions by Maxine and Ronald Linde, and Cecil and Sally Drinkward, and
the strong support of the California Institute of Technology, without
which this project would not have been possible.  In addition, this
project has benefited greatly from the computing facilities available at
CITA, and from discussions with other members of the group at CITA not
represented as authors on this paper.

\appendix

\section{Form of the Linear Estimator}\label{app:estim}

Suppose we were to construct a simple linear ``dirty'' mosaic on the sky
obtained by a linear combination of the dirty (not deconvolved) images 
of the individual fields (e.g.\ \citealt{co93}).  In the 
{\it uv} plane, this reduces to summing (integrating) up the visibilities
from each mosaic ``tile'' with some weighting function, e.g.
\begin{equation} \label{eq:linest}
   \Delta_i = \sum_{k} Q_{ik}\,V_{k} 
\end{equation}
where for the time being we ignore the contribution from the complex
conjugates of the visibilities (see below). For illustrative purposes, let us
consider only a single frequency channel and write the estimator as a 
function $\Delta(\bvec{u})$, where $\Delta_i=\Delta(\bvec{u}_i)$,
which in the absence of instrumental noise is given by
\begin{equation} \label{eq:estcont}
   \Delta(\bvec{u}) = \int d^2\bvec{v}\,\int d^2\bvec{x}\,
   {\cal F}(\bvec{x},\bvec{v})\,{\cal Q}(\bvec{u},\bvec{x},\bvec{v})\,
   \langle V(\bvec{x},\bvec{v}) \rangle
\end{equation}
with kernel ${\cal Q}$, sky and aperture plane sampling given
by ${\cal F}$, and where
\begin{equation} \label{eq:viscont}
   V(\bvec{x},\bvec{v}) = 
   \int d^2\bvec{v}'\,\tilde{I}(\bvec{v}')\,\tilde{A}(\bvec{v}-\bvec{v}')\,
      e^{2\pi i \sbvec{v}'\cdot\sbvec{x}}
\end{equation}
is the visibility at pointing position $\bvec{x}$ and {\it uv} locus
$\bvec{v}$ from equation (\ref{eq:visi}).  In practice, the sampling function
is just a series of delta functions
\begin{equation} \label{eq:fsamp}
   {\cal F}(\bvec{x},\bvec{v}) = \sum_k \omega_k\,
      \delta^2(\bvec{x}-\bvec{x}_k)\,\delta^2(\bvec{v}-\bvec{u}_k)
\end{equation}
over the measured visibilities $k=1,\ldots,N_{\rm vis}$ each with weight
$\omega_k$.

As an ansatz, we let the mosaicing kernel have the form
\begin{equation}
   {\cal Q}(\bvec{u},\bvec{x},\bvec{v}) =  Q(\bvec{v}-\bvec{u})\,
   e^{-2\pi i \sbvec{u}\cdot\sbvec{x}} \label{eq:kernel}
\end{equation}
where $Q$ is the interpolating kernel.  Furthermore, let us assume that
the {\it uv}-plane coverage is the same for all mosaic pointings, and thus
${\cal F}(\bvec{x},\bvec{v})$ is separable 
\begin{equation}
   {\cal F}(\bvec{x},\bvec{v}) \equiv F(\bvec{v})\,G(\bvec{x})
\end{equation}
where $F(\bvec{v})$ and $G(\bvec{x})$ are the sampling and weighting in the
two domains.
Combining these and rearranging terms, we get 
\begin{eqnarray}
   \Delta(\bvec{u}) & = & 
      \int d^2\bvec{v}'\, \tilde{I}(\bvec{v}') \, 
      \int d^2\bvec{v}\, F(\bvec{v})\,Q(\bvec{v}-\bvec{u})\,
         \tilde{A}(\bvec{v}-\bvec{v}')\,
      \int d^2\bvec{x}\, G(\bvec{x})\,
         e^{-2\pi i (\sbvec{u}-\sbvec{v}')\cdot\sbvec{x}} 
      \label{eq:mos2a} \\
   & = & \int d^2\bvec{v}'\, 
      \tilde{I}(\bvec{v}')\,\tilde{G}(\bvec{u}-\bvec{v}')\, 
      \int d^2\bvec{v}\, F(\bvec{v})\,Q(\bvec{v}-\bvec{u})\,
         \tilde{A}(\bvec{v}-\bvec{v}')
      \label{eq:mos2} 
\end{eqnarray}
where in equation (\ref{eq:mos2}) we used the fact that the final right-hand
side integral in equation (\ref{eq:mos2a}) is the Fourier transform
$\tilde{G}$ of the mosaic function ${G}$.

For an infinite continuous mosaic,
$\tilde{G}(\bvec{v}'-\bvec{u})=\delta^2(\bvec{v}'-\bvec{u})$ and thus
\begin{equation} \label{eq:mos3} 
   \Delta(\bvec{u}) = \tilde{I}(\bvec{u})\, 
      \int d^2\bvec{v}\, F(\bvec{v})\,Q(\bvec{v}-\bvec{u})\,
         \tilde{A}(\bvec{v}-\bvec{u}).
\end{equation}
If we wish to recover 
$\Delta(\bvec{u}) = \tilde{I}(\bvec{u})$
in this limit, then
\begin{equation} \label{eq:qvu}
   Q(\bvec{v}-\bvec{u}) =  
      {1\over z(\bvec{u})}\,\tilde{A}^\ast(\bvec{v}-\bvec{u})
\end{equation}
with normalization
\begin{equation}
   z(\bvec{u}) = \int d^2\bvec{v}\, F(\bvec{v})\,
      \tilde{A}^2(\bvec{v}-\bvec{u})
   \label{eq:zu}
\end{equation}
will fulfill our requirements.  We have chosen 
$\tilde{A}^\ast(\bvec{v}-\bvec{u})$ as the {\it uv} kernel as it 
reproduces the least-squares estimate of the sky brightness in the
linear mosaic \citep{co93}.  Then, equation (\ref{eq:mos2}) becomes
\begin{equation} \label{eq:mos4} 
   \Delta(\bvec{u}) = 
      {1\over z(\bvec{u})}\,\int d^2\bvec{v}'\, 
      \tilde{I}(\bvec{v}')\,\tilde{G}(\bvec{u}-\bvec{v}')\, 
      \int d^2\bvec{v}\, F(\bvec{v})\,\tilde{A}^\ast(\bvec{v}-\bvec{u})\,
      \tilde{A}(\bvec{v}-\bvec{v}')
\end{equation}
which has a width controlled by the narrower of the width of $\tilde{A}^2$ or
the width of $\tilde{G}$.  Thus, by widening the mosaic $G({\bf x})$ to a
larger area than the beam $A$, we will fill in the desired information inside 
the $\tilde{A}$ smeared patches in the {\it uv}-plane.  Thus, a properly
sampled $M^2$ mosaic will fill in a $M^2$ sub-grid
within each {\it uv} cell you would have normally had for a single pointing,
and thus an $M^2$ mosaic consisting of $N^2$ ``images'' each is equivalent to
a {\it uv} super-grid of size $(M \times N)^2$ (e.g.\ \citealt{ekers}).

Note that for a non-continuous mosaic, there will be ``aliases'' in the {\it
uv} plane separated by the inverse of the mosaic spacing in the sky
\citep{co88}.  Ideally, we would like the separation between aliased copies to
be larger than the extent of the beam transform, which is satisfied for
$\Delta x \leq \lambda / 2D$
which for $D=90$~cm corresponds to $21\farcm6$ at $26.5$~GHz and 
only $16\farcm1$ at $35.5$~GHz, the centers of the extremal CBI bands.
The spacing used in the CBI mosaics is a compromise between the aliasing
limits over the bands and the desire to have a fewer number of pointings
on a convenient grid.  We chose to observe with pointing centers separated
by $20^\prime$, which is sub-optimal above $27.5$~GHz.  However, the effect
of aliasing is small, with the overlap point $a^{-1} - D\,\lambda^{-1}$
occurring at the $0.61\%$ point of $\tilde{A}$ at 31~GHz, and the $6.5\%$ 
point for the highest frequency channel at $35.5$~GHz.

We obtain the gridding kernel $Q_{ik}$ of equation (\ref{eq:linest}) 
corresponding to equation (\ref{eq:qvu}) by using
the discrete sampling in equation (\ref{eq:fsamp})
\begin{equation} \label{eq:moskern} 
   Q_{ik} = {\omega_k \over z_i}\,\tilde{A}_k^\ast(\bvec{u}_k-\bvec{u}_i)
     \,e^{-2\pi i \sbvec{u}_i\cdot\sbvec{x}_k} 
\end{equation}
with visibility weights $\omega_k$, and normalization factor $z_i$.  
The discrete form of the normalization derived in equation (\ref{eq:zu}) is
\begin{equation} \label{eq:zi}
   z_i = \sum_k \omega_k \, \tilde{A}_k^2(\bvec{u}_k-\bvec{u}_i).
\end{equation}
Then,
\begin{equation} \label{eq:linest2}
   \Delta_i = {1 \over z_i}\,\sum_{k} \omega_k\,
   \tilde{A}_k^\ast(\bvec{u}_k-\bvec{u}_i)\,V_{k}\,
   e^{-2\pi i \sbvec{u}_i\cdot\sbvec{x}_k} 
\end{equation}
is the weighted sum of visibilities used for the estimators.  Note that
because $V(\bvec{u})=V^\ast(-\bvec{u})$, there are also visibilities
$V_{k'}$ for which $-\bvec{u}_{k'}$ lies within the support range around 
$\bvec{u}_i$, i.e.\
$\big|\tilde{A}_{k'}^\ast(-\bvec{u}_{k'}-\bvec{u}_i)\big|>0$.  Thus,
we should add in the complex conjugates $V^\ast_{k}$
\begin{equation} \label{eq:linest3}
   \Delta_i = {1 \over z_i}\,\sum_{k} \omega_k\,
   \big\{ \tilde{A}_k^\ast(\bvec{u}_k-\bvec{u}_i)\,V_{k} + 
      \tilde{A}_k^\ast(-\bvec{u}_k-\bvec{u}_i)\,V^\ast_{k} \big\}
      \,e^{-2\pi i \sbvec{u}_i\cdot\sbvec{x}_k}.
\end{equation}
To do this, we construct another kernel $\overline{Q}_{ik}$ 
\begin{equation} \label{eq:moskernalt} 
   \overline{Q}_{ik} = {\omega_k \over z_i}\, 
   \tilde{A}_k^\ast(-\bvec{u}_k-\bvec{u}_i)\,
   e^{-2\pi i \sbvec{u}_i\cdot\sbvec{x}_k}
\end{equation}
which will gather the appropriate $V^\ast_{k}$, giving
\begin{equation} \label{eq:linoptot}
   \Delta_i = \sum_k \big\{ Q_{ik}\,V_k + \overline{Q}_{ik}\,V^\ast_k \big\}
\end{equation}
for the final form of our linear estimator.

For estimated visibility variances $\epsilon^2_k$, the optimal weighting 
factor (in the least-squares estimation sense) is given by
\begin{equation} \label{eq:viswt}
   \omega_k = { 1 \over \epsilon^{2}_k }
\end{equation}
but may also include factors based on position in the mosaic or 
frequency channel.  For example, up until now we have neglected the frequency
dependence of the observed visibilities.  
If we are reconstructing an intensity field with a uniform
flux density spectrum, then no changes need be made.  If there is a frequency
dependence, such as that for a power-law foreground (e.g.\ eq.[\ref{eq:alpha}])
or the thermal spectrum of the CMB (e.g.\ eq.[\ref{eq:fplanck}]), then
the visibilities should be scaled and weighted by the appropriate
factor $f_k$ when gridded in order to properly estimate 
$\tilde{I}_0(\bvec{u}_k)$ or $\tilde{T}(\bvec{u}_k$
respectively.  For example, for the CMB using equation (\ref{eq:fplanck}) 
for the spectrum, we find
\begin{eqnarray} 
   Q^{\rm T}_{ik} & = & {f_{\rm T}^{-1}(\nu_k)\,\omega_k \over z_i}\, 
   \tilde{A}_k^\ast(\bvec{u}_k-\bvec{u}_i) \nonumber \\
   \omega_k & = & { f_{\rm T}^2(\nu_k) \over \epsilon^{2}_k }.
   \label{eq:qtherm} 
\end{eqnarray}
In practice for the CBI, the frequency
range of the data is not great enough for the spectral weighting factor
to matter, and we therefore use the default weighting given in equation
(\ref{eq:viswt}).  This will therefore be slightly non-optimal in the
signal-to-noise sense (it will not be the minimum-variance estimator) but
it will not introduce a bias in the bandpowers.

The choice of the normalization $z_i$ is somewhat arbitrary, as it only
determines the units of the $\Delta_i$ and not the correlation properties.
However, this can be important if we wish to use the estimators to make images
using the formalism of \S\ref{sec:image}.
For instance, the normalization given in equation (\ref{eq:zi}) has the 
drawback of diverging in cells where
all the $\tilde{A}_k^2$ are vanishingly small (such as the innermost and
outermost supported parts of the {\it uv}-plane), and will produce images
with heightened noise on short and long spatial wavelengths. It is 
therefore more convenient to use the alternate normalization
\begin{equation} \label{eq:zk1}
   z_i = \sum_k \omega_k \, \tilde{A}^\ast_k(\bvec{u}_k-\bvec{u}_i)
\end{equation}
which when inserted into equation (\ref{eq:linest2}) will properly
normalize the weighted sums of visibilities.  This will then produce 
images with the desired units of Janskys per beam (see
\S\ref{sec:image}).   We therefore use equation (\ref{eq:zk1}) for the
normalization in our CBI pipeline.

\section{Source Counts and the Residual Covariance Matrix}\label{app:counts}

We wish to calculate ${C}^{\rm res}_\nu$ (cf.\ eq.[\ref{eq:creskk}]) using
equation (\ref{eq:resampint}) with $\nu_k=\nu_{k'}=\nu$.
If $p(\alpha|\nu_0)$ is independent of 
flux density, then
\begin{equation} \label{eq:resampdef}
  {C}^{\rm res}_\nu = 
    \int_{-\infty}^{\infty} d\alpha\,p(\alpha|\nu_0)\,
    \left( {\nu \over \nu_0} \right)^{2\alpha}
    \int_0^{S_{\rm max}(\alpha)} dS\,S^2\,{ dN(S) \over dS }\,
\end{equation}
where we have left in the possibility that the upper flux density cutoff will
depend on spectral index (see below) and set the lower flux density cutoff to
zero (the results for realistic power-law counts with $\gamma>-2$ 
are insensitive to the lower cutoff, but one can easily be included).
As an example for the calculation of the fluctuation power due to residual
sources in the Gaussian limit, consider power-law integral source counts
\begin{equation}
   N(>S) = N_0\,\left( {S \over S_0} \right)^\gamma  \Rightarrow 
   { dN(S) \over dS } = -{\gamma\,N_0 \over S_0}\,
   \left( {S \over S_0} \right)^{\gamma-1}
\end{equation}
where $N(>S)$ is the mean number density of sources with flux density
{\it greater} than $S$ at frequency $\nu_0$, and a Gaussian spectral index
distribution at frequency $\nu_0$
\begin{equation}
   p(\alpha|S,\nu_0) = p(\alpha|\nu_0) = 
   {1\over \sqrt{2\pi\sigma^2_\alpha}}\,
   e^{-{(\alpha - \alpha_0)^2\over 2\sigma^2_\alpha}}.
   \label{eq:gausalpha}
\end{equation}

First consider the case where there is a fixed flux density upper cutoff
$S_{\rm max}$ at the frequency where the number counts are defined.  The
two parts of equation (\ref{eq:creskk}) separate easily,
where the source count part of the integral is
\begin{equation}
   \int_0^{S_{\rm max}} dS\,S^2\,{ dN(S) \over dS }\,
   = -{\gamma \over \gamma+2}\,N_0\,S^2_0\,
   \left( {S_{\rm max} \over S_0} \right)^{\gamma+2}.
\end{equation}
For the distribution in equation (\ref{eq:gausalpha}), 
the integral over $\alpha$ becomes
\begin{eqnarray}
 \int_{-\infty}^{\infty} d\alpha\,p(\alpha|\nu_0)\,
    \left( {\nu \over \nu_0} \right)^{2\alpha}
 & = & 
   \int_{-\infty}^{\infty} { d\alpha \over \sqrt{2\pi\sigma^2_\alpha}}\,
   e^{-{(\alpha - \alpha_0)\over 2\sigma^2_\alpha}}\,
   e^{2\alpha\,\beta} \nonumber \\
 & = & e^{-{(\bar{\alpha}^2 - \alpha^2_0)^2\over 2\sigma^2_\alpha}} \,
   \int_{-\infty}^{\infty} { d\alpha  \over \sqrt{2\pi\sigma^2_\alpha} }\,
   e^{-{(\alpha - \bar{\alpha})^2\over 2\sigma^2_\alpha}} \nonumber \\
 & = & e^{2\alpha_{\rm eff}\,\beta }
\end{eqnarray}
where $\beta = \ln(\nu/\nu_0)$, and
\begin{eqnarray}
   \alpha_{\rm eff} & = & { \bar{\alpha}^2 - \alpha^2_0 \over 
      4\beta\,\sigma^2_\alpha } = 
   \alpha_0 + \beta\,\sigma^2_\alpha
   \nonumber \\
   \bar{\alpha} & = & \alpha_0 + 2\beta\,\sigma^2_\alpha
   \label{eq:alphaeff}
\end{eqnarray}
where $\bar{\alpha}$ is the mean of the extrapolated spectral index
distribution which remains a Gaussian, and the effective spectral index
$\alpha_{\rm eff}$ for the spectral component is shifted from the mean 
spectral index of the input distribution $\alpha_0$ by the combination of the
scatter in the $\alpha$ and the lever arm $\beta$ from the frequency
extrapolation.  Putting these together, we get
\begin{equation}
   {C}^{\rm res}_\nu = -{\gamma \over \gamma+2}\,N_0\,S^2_0\,
   \left( {S_{\rm max} \over S_0} \right)^{\gamma+2}
   \,e^{2\alpha_{\rm eff}\,\beta }.
\end{equation}


One can also deal with the case where there is an upper flux density
cutoff $\hat{S}_{\rm max}$ imposed at a frequency $\hat{\nu}$ other than 
$\nu_0$ where the $N(S)$ distribution is defined.  In this case, the
flux density cutoff in equation (\ref{eq:resampdef}) is
\begin{equation}
  S_{\rm max}(\alpha) = \bar{S}_{\rm max}\,e^{(\alpha_0-\alpha)\,\hat{\beta}}
  \hspace{2cm} \bar{S}_{\rm max} = 
    \hat{S}_{\rm max}\,e^{-\alpha_0\,\hat{\beta}} 
  \label{eq:snumax}
\end{equation}
where $\hat{\beta} = \ln(\hat{\nu}/\nu_0)$, and $\bar{S}_{\rm max}$ is the
cutoff $\hat{S}_{\rm max}$ extrapolated to $\nu_0$ using $\alpha_0$.  Then, 
\begin{equation}
   \int_{0}^{S_{\rm max}(\alpha)} dS\,S^2\,{ dN(S) \over dS } =
    -{\gamma\over \gamma+2}\,N_0\,S^2_0\,
    \left( {\bar{S}_{\rm max} \over S_0} \right)^{\gamma+2}
    \,e^{(\alpha_0-\alpha)\,\hat{\beta}\,(\gamma+2)} \nonumber \\
\end{equation}
and thus
\begin{eqnarray} \label{eq:resampcut}
  {C}^{\rm res}_\nu
  & = & -{\gamma\over \gamma+2}\,N_0\,S^2_0\,
    \left( {\bar{S}_{\rm max} \over S_0} \right)^{\gamma+2}\,
    \int_{-\infty}^{\infty} d\alpha\,p(\alpha|\nu_0)\,
    e^{2\alpha\,\beta} \,
    e^{(\alpha_0-\alpha)\,\hat{\beta}\,(\gamma+2)} \nonumber \\
  & = & -{\gamma\over \gamma+2}\,N_0\,S^2_0\,
    \left( {\bar{S}_{\rm max} \over S_0} \right)^{\gamma+2}\,
    e^{2\hat{\alpha}_{\rm eff}\,\beta}
\end{eqnarray}
with
\begin{eqnarray}
  \hat{\alpha}_{\rm eff} & = & \alpha_0 + \beta\,\sigma^2_\alpha\,\kappa^2  
  \nonumber \\
  \bar{\alpha} & = & \alpha_0 + 2\beta\,\sigma^2_\alpha\,\kappa
  \nonumber \\
  \kappa & = & 1 - {\hat{\beta} \over 2\beta}\,(\gamma+2)
  \label{eq:alphaeffhat}
\end{eqnarray}
where $\kappa$ gives the modification of the effective spectral index
due to the change in the frequency at which the cutoff is done.


One often has an upper flux density cutoff at two different
frequencies.  For example, sources that are extrapolated to be bright at the
CMB observing frequency will have been detected and subtracted.  If there
is a flux density cutoff of $\hat{S}_{\rm max}$ imposed at a frequency
$\hat{\nu}$ as before, but an additional upper cutoff of
$\hat{S}^\prime_{\rm max}$ at another frequency $\hat{\nu}^\prime$, then there
is a critical spectral index
\begin{equation} \label{eq:alpcrit}
  \alpha_{\rm crit} = { \ln( \hat{S}^\prime_{\rm max} / \hat{S}_{\rm max} )
   \over \ln( \hat{\nu}^\prime / \hat{\nu} ) }
\end{equation}
above which the effective cutoff $\bar{S}_{\rm max}$ of equation 
(\ref{eq:snumax}) changes from that 
appropriate to $\hat{\nu}$ to that at $\hat{\nu}^\prime$ (assuming
$\hat{\nu}^\prime > \hat{\nu}$).  Thus, the integral over $\alpha$ in
equation (\ref{eq:resampcut}) will be broken into two pieces,
\begin{eqnarray}
  J_1 & = & -{\gamma\over \gamma+2}\,N_0\,S^2_0\,
    \left( {\bar{S}_{\rm max} \over S_0} \right)^{\gamma+2}\,
    e^{2\hat{\alpha}_{\rm eff}\,\beta}\,
    \int_{-\infty}^{\alpha_{\rm crit}} 
    {d\alpha \over \sqrt{2\pi\sigma^2_\alpha}}
    \,e^{-{(\alpha-\bar{\alpha})^2\over 2\sigma^2_\alpha} } 
    \nonumber \\
  J_2 & = & -{\gamma\over \gamma+2}\,N_0\,S^2_0\,
    \left( {\bar{S}^\prime_{\rm max} \over S_0} \right)^{\gamma+2}\,
    e^{2\hat{\alpha}^\prime_{\rm eff}\,\beta}\,
    \int_{\alpha_{\rm crit}}^{\infty} 
    {d\alpha \over \sqrt{2\pi\sigma^2_\alpha}}
    \,e^{-{(\alpha-\bar{\alpha}^\prime)^2\over 2\sigma^2_\alpha} }
    \label{eq:intj}
\end{eqnarray}
where ${C}^{\rm res}_\nu = J_1 + J_2$.
The quantities in $J_1$ are as defined in equations (\ref{eq:snumax})
through (\ref{eq:alphaeffhat}), and the parameters in $J_2$ are defined
in the same way but using the higher frequency $\hat{\nu}^\prime$.  The
truncated Gaussian integrals are just
the integrated probabilities for the normal distribution
\begin{equation}
  F(x) = {1\over \sqrt{2\pi}}\int_{-\infty}^{x} dt\,e^{-{t^2\over2}}
       = {1\over2} + {1\over2}\,{\rm erf}\left( {x\over\sqrt{2}} \right)
\end{equation}
with ${\rm erf}(z)$ the Error Function.  Then,
\begin{eqnarray} \label{eq:j1}
  J_1 & = & -{\gamma\over \gamma+2}\,N_0\,S^2_0\,
    \left( {\bar{S}_{\rm max} \over S_0} \right)^{\gamma+2}\,
    e^{2\hat{\alpha}_{\rm eff}\,\beta}\,F(x_{\rm crit}) \\
  J_2 & = & -{\gamma\over \gamma+2}\,N_0\,S^2_0\,
    \left( {\bar{S}^\prime_{\rm max} \over S_0} \right)^{\gamma+2}\,
    e^{2\hat{\alpha}^\prime_{\rm eff}\,\beta}\,
    \big[ 1 - F(x^\prime_{\rm crit}) \big]
  \label{eq:j2}
\end{eqnarray}
where $x_{\rm crit}=(\alpha_{\rm crit}-\bar{\alpha})/\sigma_\alpha$ and
$x^\prime_{\rm crit}=(\alpha_{\rm crit}-\bar{\alpha}^\prime)/\sigma_\alpha$.

As an example, consider the source counts presented in
Paper~II (\S4.3.2), with $N_0= 9.2\times10^3\,{\rm sr}^{-1}$ above 
$S_0=10$~mJy at $\nu_0=31$~GHz and $\gamma=-0.875$, which gives
\begin{equation}
   -{\gamma \over \gamma+2}\,N_0\,S^2_0 = 0.715\,{\rm Jy}^2\,{\rm sr}^{-1}
\end{equation}
as the raw source power. In the analysis described there, Mason et al. find
that a Gaussian 1.4~GHz to 31~GHz spectral index distribution with
$\alpha_0=-0.45$ and $\sigma_\alpha=0.37$ fits the observed data.
The CBI and OVRO direct measurements have 
a cutoff of $S_{\rm max}=25$~mJy at 31~GHz (sources brighter than this have
been subtracted from the CBI data and have residual uncertainties
placed in a source covariance matrix), and sources above 
$\hat{S}_{\rm max}=3.4$~mJy at $\hat{\nu}=1.4$~GHz have already been accounted
for in a second source matrix.  Therefore, the critical spectral index is
$\alpha_{\rm crit}=0.644$ from equation (\ref{eq:alpcrit}).  For 
$\alpha\geq\alpha_{\rm crit}$, the 31~GHz cutoff holds.  Since the cutoff
and source distribution are at the same frequency as the observations
$\nu=\nu_0$, there is no extrapolation factor
$\beta=0$ and the spectral index distribution is unchanged
($\bar{\alpha}^\prime=\alpha_0$). Then, $x^\prime_{\rm crit}=2.957$ and 
$1-F(x^\prime_{\rm crit})\approx1.56\times10^{-3}$, so
\begin{equation}
   J_2 = -{\gamma \over \gamma+2}\,N_0\,S^2_0\,
         \left( {S_{\rm max} \over S_0} \right)^{\gamma+2}\,
         \big[ 1 - F(x^\prime_{\rm crit}) \big] 
       = 0.003\,{\rm Jy}^2\,{\rm sr}^{-1}
\end{equation}
for the flat-spectrum tail of the spectral index integral.  
The rest of the integral uses the 1.4~GHz
cutoff, which we extrapolate using the mean spectrum to 31~GHz using
equation (\ref{eq:snumax}), 
\begin{equation}
  \bar{S}_{\rm max} = \hat{S}_{\rm max}\,e^{-\alpha_0\,\hat{\beta}} 
  = 0.843\,{\rm mJy},
  \hspace{2cm} \hat{\beta}=-3.098.
\end{equation}
Because $\beta=0$, we have to modify the quantities in
equation (\ref{eq:alphaeffhat}) by explicitly expanding the terms in $\kappa$,
and canceling remaining terms in $\beta$, giving
\begin{equation}
  \bar{\alpha} = \alpha_0 - \hat{\beta}\,\sigma^2_\alpha\,(\gamma+2)
     = 0.027,
  \hspace{2cm}
  2\hat{\alpha}_{\rm eff}\,\beta =
     {1\over2}\,\hat{\beta}^2\,\sigma^2_\alpha\,(\gamma+2)^2
     = 0.831
\end{equation}
which can then be inserted into equation (\ref{eq:j1}), giving
\begin{equation} 
  J_1 = -{\gamma \over \gamma+2}\,N_0\,S^2_0\,
    \left( {\bar{S}_{\rm max} \over S_0} \right)^{\gamma+2}\,
    e^{2\hat{\alpha}_{\rm eff}\,\beta}\,F(x_{\rm crit})
      = 0.100\,{\rm Jy}^2\,{\rm sr}^{-1}
\end{equation}
for $x_{\rm crit}=1.667$, $F(x_{\rm crit})\approx0.952$, and thus we expect
\begin{equation}
   C^{\rm res}_\nu = 0.10\, {\rm Jy}^2\,{\rm sr}^{-1}
\end{equation}
for the amplitude of the residual sources in the CBI fields.  In Paper~II,
it is noted that there is a 25\% uncertainty on $N_0$, and more importantly
the power-law slope of the source counts could conceivably
be as steep as $\gamma=-1$.  Taking the extreme of $\gamma=-1$, we get
\begin{equation}
   C^{\rm res}_\nu = 0.15\, {\rm Jy}^2\,{\rm sr}^{-1}
\end{equation}
using the above procedure.  We thus conservatively estimate a 50\% uncertainty
on the value of $C^{\rm res}_\nu$ derived in this manner.  Note that in
Paper~II we actually use the value of 
$C^{\rm res}_\nu=0.08\, {\rm Jy}^2\,{\rm sr}^{-1}$  
derived using a Monte-Carlo procedure emulating the integrals in equation
(\ref{eq:intj}) but using the actual observed distribution of source flux
densities and spectral indices.  The agreement between these two estimates
shows the efficacy of this procedure in practice.

\section{Comparison With the HM Method} \label{app:hobson}

Recently, \citet[HM]{hob02} have independently proposed a binned {\it uv}-plane
method that is somewhat similar to ours, though it is more directly related
to the ``optimal maps'' of \citet{bc01}.  HM use a gathering mapping ${\bf H}$
(${\bf M}$ in their notation) 
\begin{equation}
  \bvec{V} = {\bf H}\,\bvec{s} + \bvec{e}
\end{equation}
rather than our scattering kernel ${\bf Q}$ of equation (\ref{eq:linop}).
In the HM method, the vector $\bvec{s}$ can be thought of as a set of
ideal pixels in the {\it uv}-plane.  They show that the likelihood depends
upon binned visibilities $\bvec{v}$ and noise $\bvec{n}$
\begin{eqnarray}
  \bvec{v} & = & ({\bf H}^\dag\,{\bf E}^{-1}\,{\bf H})^{-1}
     \,{\bf H}^\dag\,{\bf E}^{-1}\,\bvec{V} = \bvec{s} + \bvec{n}
     \nonumber \\
  \bvec{n} & = & ({\bf H}^\dag\,{\bf E}^{-1}\,{\bf H})^{-1}
     \,{\bf H}^\dag\,{\bf E}^{-1}\,\bvec{e}
\end{eqnarray}
where 
\begin{equation}
  {\bf C}^{\rm N} = \langle \bvec{n}\,\bvec{n}^\dag \rangle 
  = ({\bf H}^\dag\,{\bf E}^{-1}\,{\bf H})^{-1}.
\end{equation}
The HM kernel $H_{jk}$ is chosen to equal 1 if the $\bvec{u}_j$ of
visibility $V_j$ lies in cell $k$, though other more complicated kernels
could be imagined.  The HM method will also give a calculational speedup
through the reduction in number of independent gridded estimators, and 
the use of the method is demonstrated using simulated VSA data in 
their paper.


\clearpage

\clearpage

\clearpage

\clearpage

\clearpage

\clearpage

\clearpage

\end{document}